\documentclass[10pt,journal,compsoc]{IEEEtran}

\usepackage{cite}
\usepackage{amsmath,amssymb,amsfonts}
\usepackage{algorithm}
\usepackage{algorithmic}
\usepackage{graphicx}
\usepackage{textcomp}

\def\0{{\bf 0}}
\def\1{{\bf 1}}

\def\etal{{\em et al.}}
\def\eg{{\em e.g.}}

\def\etal{{\em et al.\/}\,}

\usepackage{amsmath}
\usepackage{amssymb}
\usepackage{multirow}
\usepackage{array}
\usepackage{booktabs}
\usepackage{multirow}
\usepackage{bm}
\usepackage{graphicx}
\usepackage{epstopdf}
\usepackage{subfigure}
\usepackage{makecell}
\usepackage{cite}
\usepackage{multirow}
\usepackage{CJK}
\usepackage{soul}
\usepackage{url}
\usepackage[pagebackref=false,breaklinks=true,letterpaper=true,colorlinks,bookmarks=false]{hyperref}
\def\etal{{\em et al.}}
\def\eg{{\em e.g.}}

\usepackage{hyperref}

\def\BibTeX{{\rm B\kern-.05em{\sc i\kern-.025em b}\kern-.08em
    T\kern-.1667em\lower.7ex\hbox{E}\kern-.125emX}}

\begin{document}

\title{RelA-Diffusion: Relativistic Adversarial Diffusion for Multi-Tracer PET Synthesis from Multi-Sequence MRI}
%\maketitle 
\author{Minhui~Yu, 
Yongheng Sun,
David S. Lalush,
Jason P Mihalik,
Pew-Thian Yap,
Mingxia~Liu,~\IEEEmembership{Senior Member,~IEEE}
\thanks{M.~Yu, Y.~Sun, P.-T.~Yap and M.~Liu are with the Department of Radiology and Biomedical Research Imaging Center (BRIC), University of North Carolina at Chapel Hill, Chapel Hill, NC 27599, USA. 
M.~Yu and D.~Lalush are with the Joint Department of Biomedical Engineering, University of North Carolina at Chapel Hill and North Carolina State University, Chapel Hill, NC 27599, USA. 
J.~Mihalik is with the Department of Exercise and Sport Science, UNC-CH, Chapel Hill, NC 27599, USA.
}
%\thanks{This research was supported in part by NIH grants (Nos. AG073297, AG082938, EB035160, and NS134849). A part of the data used in this work is from ADNI and AIBL. The ADNI and AIBL investigators provide data but are not involved in data processing, analysis, and writing. A comprehensive list of ADNI investigators is accessible \href{https://adni.loni.usc.edu/wp-content/uploads/how\_to\_apply/ADNI\_Acknowledgement\_List.pdf}{online}.}
}

\maketitle

\begin{abstract}
Multi-tracer positron emission tomography (PET) provides critical insights into diverse neuropathological processes such as tau accumulation, neuroinflammation, and $\beta$-amyloid deposition in the brain, making it indispensable for comprehensive neurological assessment. 
However, routine acquisition of multi-tracer PET is limited by high costs, radiation exposure, and restricted tracer availability.  
Recent efforts have explored deep learning approaches for synthesizing PET images from structural MRI. 
While some methods rely solely on T1-weighted MRI, others incorporate additional sequences such as T2-FLAIR to improve pathological sensitivity. 
However, existing methods often struggle to capture fine-grained anatomical and pathological details, resulting in artifacts and unrealistic outputs.
%To address these limitations, 
To this end, we propose RelA-Diffusion, a Relativistic Adversarial Diffusion framework for multi-tracer PET synthesis from multi-sequence MRI. 
By leveraging both T1-weighted and T2-FLAIR scans as complementary inputs, RelA-Diffusion captures richer structural information to guide image generation. 
{\color{black}To improve synthesis fidelity, we introduce a gradient-penalized relativistic adversarial loss to the intermediate clean predictions of the diffusion model. 
This loss compares real and generated images in a relative manner, encouraging the synthesis of more realistic local structures. 
Both the relativistic formulation and the gradient penalty contribute to stabilizing the training, while adversarial feedback at each diffusion timestep enables consistent refinement throughout the generation process.} 
Extensive experiments on two datasets 
demonstrate that RelA-Diffusion outperforms existing methods in both visual fidelity and quantitative metrics, 
highlighting its potential for accurate synthesis of multi-tracer PET. 
\end{abstract}


% make the title area
%\maketitle

% \IEEEdisplaynontitleabstractindextext

\IEEEpeerreviewmaketitle

\section{Introduction}

Accurate assessment and monitoring of neurodegenerative disorders rely heavily on the ability to capture diverse and subtle pathological processes within the brain. 
%Multi-tracer 
Brain positron emission tomography (PET) is a powerful imaging modality that enables the \emph{in vivo} visualization of various molecular targets, including $\beta$-amyloid deposition with $^{11}$C-PIB (PIB), tau pathology with $^{18}$F-AV1451 (TAU)~\cite{johnson2016tau,schwarz2016regional,antoni2013vivo,xia201318f}, and neuroinflammatory activity with $^{18}$F-PBR111 (PBR)~\cite{colasanti2014vivo}. 
\if false
Brain positron emission tomography (PET) serves as a powerful imaging modality for this purpose, enabling visualization of various molecular targets such as 
$\beta$-amyloid deposition with $^{11}$C-PIB (PIB), \emph{tau} pathology with $^{18}$F-AV1457 (TAU)~\cite{johnson2016tau,schwarz2016regional,antoni2013vivo,xia201318f}, 
%protein aggregations~\cite{johnson2016tau,schwarz2016regional,antoni2013vivo,xia201318f}
and neuroinflammatory activity with $^{18}$F-PBR111 (PBR)~\cite{colasanti2014vivo}. 
%The combination of different PET tracers 
These multi-tracer PET imaging, PET imaging is widely used to study \emph{Tau} pathology with $^{18}$F-T807~\cite{xia201318f}, 
% neuroinflammatory with $^{18}$F-PBR111~\cite{colasanti2014vivo}, 
% $\beta$-amyloid deposition with $^{11}$C-PIB~\cite{antoni2013vivo} 
\fi 
This multi-tracer PET imaging can provide complementary information, offering a more comprehensive understanding of disease mechanisms. 
However, the use of multi-tracer PET imaging in clinical and research settings is hindered by several practical constraints, including high scanning costs, radiation exposure, and limited access to specific tracers. 
%These limitations have motivated the development of alternative approaches that can approximate 
These limitations have motivated the development of alternative approaches that can synthesize 
PET without requiring tracer administration. 
% Public datasets rarely have multi-tracer PET from the same timepoint available.
%To mitigate these constraints, 
Recent research has focused on developing deep learning techniques to synthesize PET images from commonly accessible structural MRI, offering a non-invasive and cost-effective option. %alternative.
%}这里是不是可以加入参考文献

% To this end, recent research has explored the use of deep learning techniques to synthesize PET images from magnetic resonance imaging (MRI), which is non-invasive and widely available. While early efforts have demonstrated the feasibility of generating single-tracer PET images from structural MRI, most methods are limited by their reliance on only T1-weighted MRI inputs, which capture anatomical structure but lack sensitivity to certain pathological signals. Moreover, these models often employ direct regression techniques that fail to capture the complex, spatially varying characteristics of PET images, leading to unrealistic textures, blurred regions, or anatomical misalignments. Some more recent methods have attempted multi-tracer synthesis, but they often lack mechanisms to ensure local fidelity and structural sharpness across different tracer types.

\if false 
%\textcolor{cyan}{
Earlier methods for MRI-based PET image synthesis primarily employed GAN-based models. %architectures. 
Wei~\etal~\cite{wei2020predicting}~propose a conditional flexible self-attention GAN model to predict PET parametric maps from multi-sequence MRI,  training the GAN using a Sketcher–Refiner approach.  
Hu~\etal~\cite{hu2021bidirectional}~introduce a bidirectional GAN that learns both PET generation and inversion back to latent representations, aiming to effectively capture semantic content within the latent space. 
Zhang~\etal~\cite{zhang2022bpgan}~develop BPGAN, which utilizes a 3D multi-convolution U-Net generator with gradient profile and structural similarity index measure constraints to better preserve structural details in synthesized PET images. 
% \cite{zeng20223d}
% {\color{red}Despite their effectiveness in producing anatomically aligned and visually realistic outputs, GAN models frequently encounter optimization challenges, such as mode collapse, which can significantly limit the diversity and generalizability of generated results ~\cite{goodfellow2016nips}.} %这句话是不是最好放在这段的最后？因为下面的CycleGAN-based方法也是GAN方法也会遇到Mode Collapse问题？ 
Several studies have applied CycleGAN-based frameworks for paired MRI-to-PET synthesis. 
For instance, Pan~\etal~\cite{pan2021disease} introduce a method that translates between T1-weighted MRI and fluorodeoxyglucose-PET while integrating disease-specific feature disentanglement and diagnosis-guided supervision to enhance clinical relevance. 
Zhou~\etal~\cite{zhou2021synthesizing} explore translation between multiple PET tracers, conditioning the synthesis process on anatomical context derived from structural MRI. 
% While cycle-consistency constraints help preserve structural alignment, these models may still face challenges in capturing subtle metabolic patterns unique to PET due to the nature of adversarial training and can be susceptible to issues like mode collapse or generating artifacts.
Despite their effectiveness, % in producing anatomically aligned and visually realistic outputs, 
GAN-based approaches frequently encounter optimization challenges, such as mode collapse, which can significantly limit model performance~\cite{goodfellow2016nips}. % generalizability of generated results ~\cite{goodfellow2016nips}.
%}
\fi

Despite the potential of deep learning for MRI-to-PET synthesis, producing images that are both anatomically precise and pathologically accurate remains a formidable challenge. Earlier GAN-based methods~\cite{wei2020predicting,hu2021bidirectional,pan2021disease,zhang2022bpgan,zhou2021synthesizing,pan2018synthesizing}, while capable of generating sharp textures, frequently struggle with optimization instabilities and mode collapse, which can result in the loss of subtle, tracer-specific pathological signatures. 
Conversely, while more recent diffusion-based strategies~\cite{xie2024synthesizing,yu2024functional,zhong2025multi} offer superior distribution coverage and training stability, they are prone to producing overly smooth outputs in high-dimensional 3D medical volumes. Most existing frameworks fail to effectively balance the strict anatomical constraints of structural MRI with the high-frequency detail required to represent complex tracer uptake patterns, leading to synthesized results that may lack the fidelity needed for clinical decision-making.

\begin{figure*}[!t]
\setlength{\abovecaptionskip}{1pt}
\setlength{\belowcaptionskip}{0pt}
\setlength\belowdisplayskip{0pt}
\setlength{\abovecaptionskip}{0pt}
\centering
\includegraphics[width=1\textwidth]{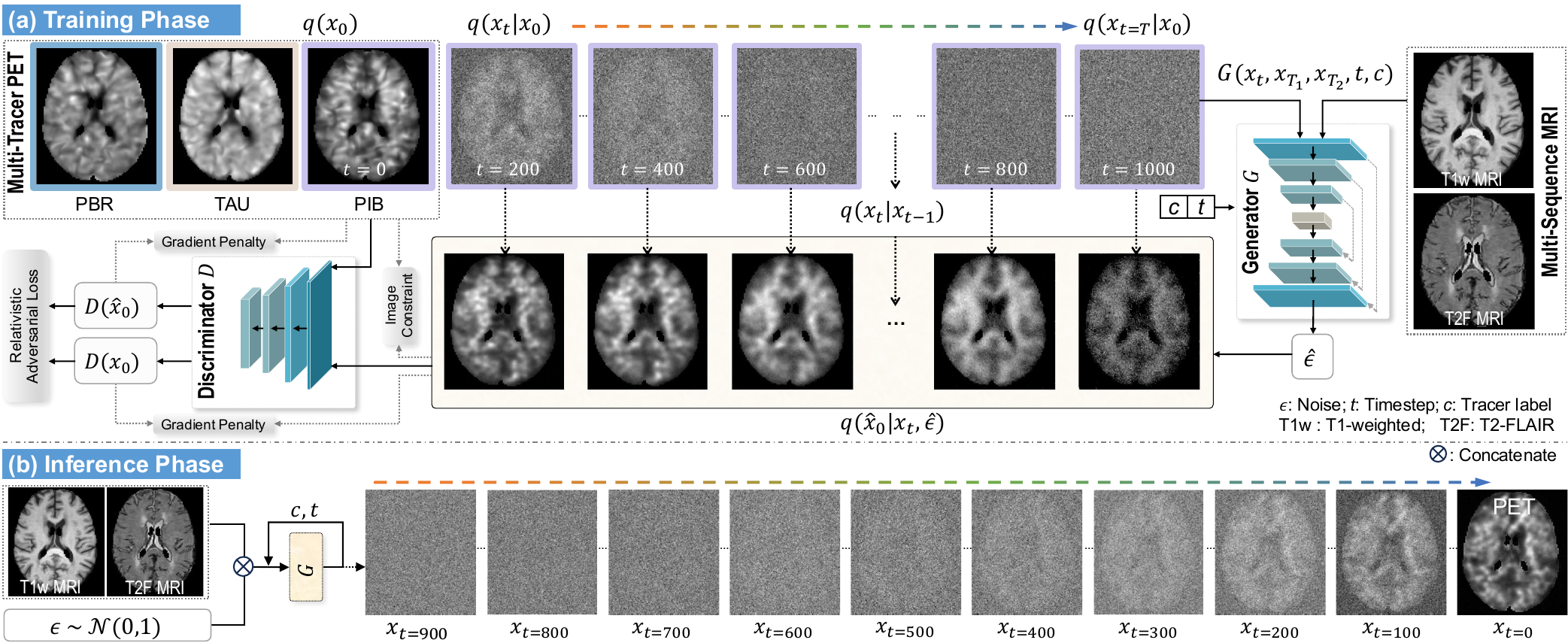}
\caption{Illustration of the proposed relativistic adversarial diffusion (RelA-Diffusion) framework that synthesizes multi-tracer PET images from multi-sequence MRI input such as T1-weighted (T1w) and T2-FLAIR (T2F) MRIs.}
\label{fig_framework}
\end{figure*}

\if false 
%\textcolor{cyan}{
More recent strategies have shifted toward diffusion %probabilistic 
models, which offer significant advantages in generating high-quality medical images. 
For example, a Joint Diffusion Attention Model (JDAM)~\cite{xie2024synthesizing} is designed to generate synthetic PET images from high-field and ultra-high-field MRI by learning the joint probability distribution between MRI input and the noisy PET output. 
Another study~\cite{yu2024functional} presents a functional imaging constrained diffusion (FICD) method, %which introduces a voxel-wise alignment loss to enforce consistency between each denoised PET prediction and its ground truth to improve quantitative fidelity. %, greatly improving quantitative fidelity and output uniqueness. 
incorporating a voxel-wise alignment loss to improve the quantitative fidelity of PET predictions by enforcing agreement with the ground truth. 
Additionally, a diffusion model MTGD~\cite{zhong2025multi} is designed to fuse multi-sequence MRI %via cross-attention 
to facilitate multi-tracer PET synthesis, showing improved performance in combining complementary MRI contrasts. 
%}

%\textcolor{cyan}{
To leverage the strengths of both paradigms, several approaches have integrated GAN objectives into diffusion frameworks. 
The study by~\cite{xiao2021tackling} models each denoising step of a diffusion process using a multi-modal conditional GAN, allowing for significantly faster sampling while maintaining high sample quality and diversity, and improving mode coverage and training stability compared to traditional GANs. 
Another study~\cite{wang2023diffusiongan} introduces Diffusion-GAN, a GAN framework that injects instance noise generated by a forward diffusion chain into the discriminator's input, thereby addressing GAN training instability and mode collapse by providing more stable and diverse learning signals. 
% While these hybrid models aim to resolve key limitations of standalone GANs and diffusion models, they often introduce complex optimization dynamics due to conflicting loss objectives, and balancing global structural coherence with local realism can remain difficult. 
While these hybrid frameworks mitigate several limitations of standalone GANs and diffusion models, they often introduce complex training dynamics due to the competing nature of adversarial and denoising objectives.
Consequently, such models may struggle to balance global structural coherence with fine-grained detail, leading to suboptimal synthesis fidelity in some cases.
% As a result, these hybrid models sometimes still produce artifacts and inconsistent boundary definition, particularly in regions with complex or subtle pathology. 
% These hybrid models aim to use adversarial feedback to sharpen images during denoising steps. 
% While they often deliver crisper outputs and enhanced sharpness compared to standalone diffusion or GAN models, they also introduce complex optimization dynamics. 
% Training becomes unstable due to conflicting loss regimes, and balancing global structural coherence with local realism remains difficult. 
% As a result, these hybrid models sometimes still produce artifacts and inconsistent boundary definition, particularly in regions with complex or subtle pathology.
%}
\fi 

%\textcolor{cyan}{
To address these critical limitations, we propose RelA-Diffusion, a novel relativistic adversarial diffusion framework specifically developed to synthesize realistic multi-tracer PET images from multi-sequence MRI scans, including T1-weighted (T1w) and T2-FLAIR (T2F) MR images. 
As illustrated in Figure~\ref{fig_framework}, RelA-Diffusion employs a conditional denoising diffusion probabilistic model that iteratively transforms Gaussian noise into realistic PET images, conditioned on multi-sequence MRI inputs. 
To improve training dynamics and enhance image fidelity, we introduce the gradient-penalized relativistic adversarial loss during training. 
This loss directly compares real and synthesized PET images to encourage sharper anatomical boundaries and more realistic local structures. 
Both the gradient penalty and the relativistic formulation help stabilize adversarial learning and regularize model behavior, reducing artifacts and promoting detailed structural representation. 
Extensive evaluations on two multi-tracer PET datasets demonstrate that RelA-Diffusion consistently outperforms state-of-the-art (SOTA) methods in terms of visual realism, anatomical accuracy, and quantitative performance metrics, establishing its effectiveness and reliability for multi-tracer PET image synthesis.
The main contributions of this work are summarized as follows:
\begin{itemize}
\vspace{-4pt}
\item We introduce RelA-Diffusion, a diffusion-based framework tailored for multi-tracer PET synthesis from multi-sequence MRI, leveraging both T1-weighted and T2-FLAIR MRI inputs for enhanced pathological sensitivity.
\item We propose to apply the gradient-penalized relativistic adversarial feedback to the intermediate clear image estimated during the training of the diffusion model, enabling consistent supervision and improved detail preservation, while stabilizing the training.
\item %Extensive experiments are conducted on two real-world datasets, with results showing that RelA-Diff achieves superior synthesis quality over existing state-of-the-art (SOTA) methods in both visual and quantitative evaluations.
Extensive experiments on two datasets suggest that RelA-Diffusion consistently outperforms existing state-of-the-art methods in visual fidelity and quantitative metrics. 
\end{itemize}

The remainder of this paper is organized as follows. Section~\ref{S_relatedWork} reviews the related work. Section~\ref{S_method} introduces the proposed RelA-Diffusion framework, while Section~\ref{S_experiment} presents the experiments and comparative analysis. In Section~\ref{S_discussion}, we discuss the impact of key components of the proposed method and present the limitations of this study alongside directions for future work. Finally, Section~\ref{S_conclusion} concludes the paper.

\section{Related Work} \label{S_relatedWork}
\subsection{GAN-based PET Image Synthesis}
\if false 
Early efforts in MRI-to-PET synthesis primarily utilized Generative Adversarial Networks (GANs) due to their ability to produce sharp, high-frequency details. 
For instance, Wei~\etal~\cite{wei2020predicting}  proposed a Sketcher-Refiner GAN that decomposed the synthesis task into anatomical sketching and content refinement. 
Similarly, Hu~\etal~\cite{hu2021bidirectional} introduced a bidirectional framework designed to capture semantic latent content through dual-path learning. Despite these successes, GANs are notoriously difficult to stabilize.  Pan \etal~\cite{pan2021disease} noted that the competing nature of the generator and discriminator often leads to mode collapse, where the model fails to represent the full variety of pathological tracer uptake, potentially leading to misleading clinical interpretations. 
\fi 

Earlier methods for MRI-based PET image synthesis primarily employed GAN-based models. %architectures. 
Wei~\etal~\cite{wei2020predicting}~propose a conditional flexible self-attention GAN model to predict PET parametric maps from multi-sequence MRI,  training the GAN using a Sketcher–Refiner approach.  
Hu~\etal~\cite{hu2021bidirectional}~introduce a bidirectional GAN that learns both PET generation and inversion back to latent representations, aiming to effectively capture semantic content within the latent space. 
Zhang~\etal~\cite{zhang2022bpgan}~develop BPGAN, which utilizes a 3D multi-convolution U-Net generator with gradient profile and structural similarity index measure constraints to better preserve structural details in synthesized PET images. 
% \cite{zeng20223d}
% {\color{red}Despite their effectiveness in producing anatomically aligned and visually realistic outputs, GAN models frequently encounter optimization challenges, such as mode collapse, which can significantly limit the diversity and generalizability of generated results ~\cite{goodfellow2016nips}.} %这句话是不是最好放在这段的最后？因为下面的CycleGAN-based方法也是GAN方法也会遇到Mode Collapse问题？ 
Several studies have applied CycleGAN-based frameworks for paired MRI-to-PET synthesis. 
For instance, Pan~\etal~\cite{pan2021disease} introduce a method that translates between T1-weighted MRI and fluorodeoxyglucose-PET while integrating disease-specific feature disentanglement and diagnosis-guided supervision to enhance clinical relevance. 
Zhou~\etal~\cite{zhou2021synthesizing} explore translation between multiple PET tracers, conditioning the synthesis process on anatomical context derived from structural MRI. 
% While cycle-consistency constraints help preserve structural alignment, these models may still face challenges in capturing subtle metabolic patterns unique to PET due to the nature of adversarial training and can be susceptible to issues like mode collapse or generating artifacts.
Despite their effectiveness, % in producing anatomically aligned and visually realistic outputs, 
GAN-based approaches frequently encounter optimization challenges, such as mode collapse, which can significantly limit model performance~\cite{goodfellow2016nips}. % generalizability of generated results ~\cite{goodfellow2016nips}.
%}

\subsection{Diffusion Models for Medical Imaging}
\if false 
Recently, Denoising Diffusion Probabilistic Models (DDPMs) have emerged as a powerful alternative due to their mathematically grounded training stability and superior mode coverage. In the context of PET synthesis, Joint Diffusion Attention Model (JDAM)~\cite{xie2024synthesizing}  demonstrated that learning the joint distribution between MRI and PET could yield more robust cross-modal mappings. Furthermore, the Functional Imaging Constrained Diffusion (FICD)~\cite{yu2024functional} method incorporated voxel-wise alignment losses to enforce strict agreement between the synthesized outputs and ground-truth uptake patterns. Additionally, a diffusion model MTGD~\cite{zhong2025multi} is designed to fuse multi-sequence MRI %via cross-attention 
to facilitate multi-tracer PET synthesis, showing improved performance in combining complementary MRI contrasts.  
While these models excel at capturing global image statistics, the iterative denoising process can sometimes lead to a smoothing effect. 
Recent hybrid efforts, such as Diffusion-GAN~\cite{wang2023diffusiongan}, attempt to address this by injecting noise into the discriminator's input, yet balancing structural coherence with local detail remains a significant challenge that our RelA-Diffusion framework aims to solve.
\fi 

More recent strategies have shifted toward diffusion models, which offer significant advantages in generating high-quality medical images. 
For example, a Joint Diffusion Attention Model (JDAM)~\cite{xie2024synthesizing} is designed to generate synthetic PET images from high-field and ultra-high-field MRI by learning the joint probability distribution between MRI input and the noisy PET output. 
Another study~\cite{yu2024functional} presents a functional imaging constrained diffusion (FICD) method, %which introduces a voxel-wise alignment loss to enforce consistency between each denoised PET prediction and its ground truth to improve quantitative fidelity. %, greatly improving quantitative fidelity and output uniqueness. 
incorporating a voxel-wise alignment loss to improve the quantitative fidelity of PET predictions by enforcing agreement with the ground truth. 
Additionally, a diffusion model MTGD~\cite{zhong2025multi} is designed to fuse multi-sequence MRI %via cross-attention 
to facilitate multi-tracer PET synthesis, showing improved performance in combining complementary MRI contrasts. 
%}

%\textcolor{cyan}{
To leverage the strengths of both paradigms, several approaches have integrated GAN objectives into diffusion frameworks. 
The study by~\cite{xiao2021tackling} models each denoising step of a diffusion process using a multi-modal conditional GAN, allowing for significantly faster sampling while maintaining high sample quality and diversity, and improving mode coverage and training stability compared to traditional GANs. 
Another study~\cite{wang2023diffusiongan} introduces Diffusion-GAN, a GAN framework that injects instance noise generated by a forward diffusion chain into the discriminator's input, thereby addressing GAN training instability and mode collapse by providing more stable and diverse learning signals. 
% While these hybrid models aim to resolve key limitations of standalone GANs and diffusion models, they often introduce complex optimization dynamics due to conflicting loss objectives, and balancing global structural coherence with local realism can remain difficult. 
While these hybrid frameworks mitigate several limitations of standalone GANs and diffusion models, they often introduce complex training dynamics due to the competing nature of adversarial and denoising objectives.
%Consequently, such models may struggle to balance global structural coherence with fine-grained detail, leading to suboptimal synthesis fidelity in some cases, a significant challenge that our RelA-Diffusion framework aims to solve.
Consequently, such models may struggle to balance global structural coherence with fine-grained detail, leading to suboptimal synthesis fidelity in some cases, a significant challenge that our RelA-Diffusion framework aims to solve.

\section{Proposed Method}
\label{S_method}
We propose \textbf{RelA-Diffusion}, a conditional diffusion framework for synthesizing high-fidelity multi-tracer PET images from multi-sequence MRI. 
Figure~\ref{fig_framework} shows that RelA-Diffusion introduces a stabilized adversarial supervision mechanism to diffusion training by leveraging a gradient-penalized relativistic discriminator to provide feedback on intermediate clean predictions \(\hat{x}_0\) at each training step. 
This feedback guides the learning of fine-grained tracer uptake patterns while promoting stable optimization dynamics throughout denoising process.

\subsection{Overall Framework} 
RelA-Diffusion consists of two main components:  a \emph{conditional denoising diffusion model} as the generator \( G \), and a \emph{relativistic discriminator} \( D \) used only during training. 
As illustrated in Figure~\ref{fig_framework}, the generator predicts the noise component \(\hat{\epsilon} = G(x_t, x_{T_1}, x_{T_2}, t, c)\) at each denoising step \( t \), where \(x_t\) is the noisy image, \(x_{T_1}\) and \(x_{T_2}\) are the T1w and T2F MRI inputs, and \(c\) denotes the tracer label. 
%后面T1-weighted简写成T1w, T2-FlARI简写成T2F
The clean image estimate \(\hat{x}_0\) is then computed from the predicted noise and \(x_t\) using the DDPM formulation~\cite{ho2020denoising}. 
The discriminator \( D(\hat{x}_0, x_0) \) compares the predicted \(\hat{x}_0\) with the ground-truth PET image \(x_0\) in a relativistic manner to provide adversarial feedback on the relative realism of the intermediate prediction. 
At inference time, only the trained generator \( G \) is used to synthesize PET images through the learned reverse diffusion process, without involving the discriminator.

%\subsubsection{Diffusion Process}
Our generative model follows a conditional denoising diffusion framework~\cite{ho2020denoising}, which defines a forward noising process as a fixed Markov chain that progressively corrupts a clean PET image \(x_0\) into a sequence of noisy variables \(\{x_t\}_{t=1}^T\) over \(T\) timesteps. 
{\color{black}This is achieved by adding Gaussian noise at each step according to a predefined noise schedule~\cite{ho2020denoising}.}  
The reverse denoising process is then learned to invert this degradation, reconstructing the original data by estimating the noise added at each step.

\subsubsection{Forward Diffusion Process}
Given a real PET image $x_0 \sim q(x_0)$, the forward process adds Gaussian noise over $T$ timesteps, creating a series of noisy samples $\{x_1, \cdots, x_T\}$. 
This process is defined by a fixed Markov chain:
\begin{align}
q(x_{1:T} | x_0) &= \prod\nolimits_{t=1}^T q(x_t|x_{t-1}), \label{eq:forward_process_product} \\
q(x_t|x_{t-1}) &= \mathcal{N}(x_t; \sqrt{1-\beta_t}x_{t-1}, \beta_t \mathbf{I}), 
\label{eq:forward_process_step}
\end{align}
where \(\mathcal{N}\) is Gaussian distribution with mean \(\mu_t\) and variance \(\sigma_t\), 
$\beta_t$ is a time-dependent hyperparameter that controls the noise level at each timestep, and $\mathbf{I}$ is the identity matrix indicating isotropic variance. 
%Accordingly, 
The marginal distribution at timestep $t$ can be expressed as:
\begin{align}
q(x_t|x_0) = \mathcal{N}(x_t; \sqrt{\bar{\alpha}_t}x_0, (1-\bar{\alpha}_t)\mathbf{I}),
\label{eq:forward_marginal}
\end{align}
where ${\alpha_t=1-\beta_t}$ and $\bar{\alpha}_t = \prod_{s=1}^t \alpha_s$.

%\subsection{Reverse Denoising Process.}
\subsubsection{Reverse Denoising Process}
Starting from a noisy image $x_t$, the reverse denoising process aims to progressively estimate and remove noise to recover the cleaner image $x_{t-1}$ in the previous timestep, and ultimately aims to reconstruct the original data $x_0$.
The trainable component for this process is a generator $G$ that predicts the noise $\epsilon$ added to $x_0$ during the forward diffusion process to obtain $x_t$, denoted as $\hat{\epsilon}$. 
As illustrated in the top right of Figure~\ref{fig_framework}, the generator is conditioned on multi-sequence MRI inputs (i.e., $x_{{T_1}}$ for T1w MRI and $x_{{T_2}}$ for T2F MRI) via input channel concatenation, and conditioned on the tracer label $c$ and timestep $t$ via cross-attention mechanisms, allowing the model to leverage anatomical %and pathological 
information from MRI and the target tracer type to guide PET image synthesis.
%In the \emph{training phase}, a randomly selected timestep $t$ 

In the \emph{training phase}, a randomly selected timestep \( t \in \{1, \cdots, T\} \) is sampled uniformly, and a noisy image \( x_t \) is generated by adding Gaussian noise to the ground-truth PET image \( x_0 \) according to the {\color{black} forward diffusion process:
\begin{align}
x_t = \sqrt{\bar{\alpha}_t} x_0 + \sqrt{1 - \bar{\alpha}_t} \, \epsilon, \quad \epsilon \sim \mathcal{N}(0, \mathbf{I}).
\end{align}}
The generator \( G \) is then trained to predict the added noise \( \epsilon \) from the noisy image \( x_t \). 
The training objective is to minimize the mean squared error (MSE) between the predicted noise \( \hat{\epsilon} \) and the ground-truth noise \( \epsilon \), defined as:
\begin{align}
% \mathcal{L}_{N} = \left\|
% \epsilon - \hat{\epsilon}(x_t, t, x_{{T_1}}, x_{{T_2}})
% \right\|^2.
\mathcal{L}_{N} = \frac{1}{n} \sum\nolimits_{i=1}^{n} ( \epsilon - \hat{\epsilon}(x_t, t, x_{{T_1}}, x_{{T_2}}) )^2,
\end{align}
where $n$ denotes the number of samples. 
Given this predicted noise, % \( \hat{\epsilon}(x_t, t) \), 
an estimate of the clean image $\hat{x}_0$ at timestep \( t \) can be computed as:
\begin{align}
\hat{x}_0(x_t, t) = \frac{x_t - \sqrt{1 - \bar{\alpha}_t} \, \hat{\epsilon}}{\sqrt{\bar{\alpha}_t}}.
\label{eq:x0_estimation}
\end{align}
In the \emph{inference phase}, the generator iteratively applies its noise predictions to progressively denoise an initial pure Gaussian noise input, starting from $x_T$ and moving step-by-step towards $x_0$. 
At each step, the generator predicts the noise $\hat{\epsilon}$, which is then utilized to estimate a cleaner version:
\begin{align}
x_{t-1} = \frac{1}{\sqrt{\alpha_t}} \left( x_t - \frac{1-\alpha_t}{\sqrt{1-\bar{\alpha}_t}} \hat{\epsilon} \right) + \sigma_t z,
\label{eq:x_t-1}
\end{align}
where $\sigma_t$ is the standard deviation of added noise and $z\sim \mathcal{N}(0,1)$.
This iterative sampling procedure, guided by the estimated $\hat{x}_0$ %(as defined in the Overall Framework and further detailed in Section~\ref{sec:supervision_on_x0}) 
and the conditional multi-sequence MRI and tracer label inputs, generates the multi-tracer PET image.

\subsection{Supervision on Intermediate Clean Predictions}
\label{sec:supervision_on_x0}
%In RelA-Diffusion, 
To enhance output fidelity, we introduce the following two constraints on the intermediate clean image $\hat{x}_0$ by comparing them against the true $x_0$ in RelA-Diffusion. 

\subsubsection{Image Constraint}
% (1) \textbf{\emph{Image Constraint}}:
The image-level constraint %includes an $l_1$ loss 
aims to minimize the $l_1$ loss between $\hat{x}_0$ and $x_0$, formulated as: 
\begin{align}
\mathcal{L}_{I} = \frac{1}{n} \sum\nolimits_{i=1}^{n} | x_0 - \hat{x}_0 |,
\label{eq:l1}
\end{align}
where $n$ is the number of samples in the training batch. 
This design explicitly guides $\hat{x}_0$ during training to promote more reliable outputs across the entire reverse process. 
It is essential since denoising becomes increasingly difficult at larger timesteps, where the input $x_t$ is heavily corrupted. 
Supervising $\hat{x}_0$ at all timesteps implicitly places stronger regularization on the model's behavior in high-noise diffusion steps, enabling the network to recover meaningful details even from severely degraded inputs.
This effect has been theoretically discussed in prior works~\cite{hang2023efficient,yu2024functional}
and provides a foundation for the subsequent adversarial feedback, ensuring that the discriminator receives informative and reliable intermediate predictions at all diffusion steps.

\subsubsection{Gradient-Penalized Relativistic Adversarial Constraint}
To further enhance realism %and detail 
of synthesized PET images, we integrate a \emph{relativistic adversarial loss}~\cite{jolicoeur2018relativistic,huang2024gan}. 
Unlike traditional GANs where the discriminator judges whether an image is real or fake in an absolute manner, a relativistic discriminator assesses whether a generated image is more realistic than a real image, or vice versa. 
%Traditional GAN discriminators judge images as real or fake in isolation; however, a relativistic discriminator evaluates whether a generated image is more realistic than a real one, or vice versa. 
The discriminator \( D \) is a classifier that receives both intermediate clean predictions \( \hat{x}_0 \) produced from noisy PET and the corresponding real PET image \( x_0 \). 
We perform pairwise comparison to estimate the relative authenticity between real and generated samples. 
Denote $\mathbb{P}$ and $\mathbb{Q}$ as the probability distributions of real and synthesized PET images, respectively. 
The relativistic adversarial (RA) loss is formulated as:
\begin{equation}
\small
\begin{aligned}
\mathcal{L}_R = & \mathbb{E}_{\substack{(x_0,\hat{x}_0)  \sim( \mathbb{P},\mathbb{Q})}} \left[ %\text{softplus}
f (D(\hat{x}_0)-D(x_0)) \right] \\
%\mathcal{L}_G 
&+ \mathbb{E}_{\substack{(x_0,\hat{x}_0)  \sim( \mathbb{P},\mathbb{Q})}} \left[ %\text{softplus}
f(D(x_0)-D(\hat{x}_0)) \right], 
\end{aligned}
\label{eq_RelativisticLoss}
\end{equation}
where \( D(x) \) denotes the discriminator's scalar output for image \( x \), \( D(\hat{x}) \) denotes the output for the synthesized image \( \hat{x} \), $f(y)=-\text{log}(1+e^{-y})$ is the activation function~\cite{nowozin2016f}, %添加softplus的文献 
and the expectations are computed over the batch of real and synthesized images. 
{\color{black}The first term in Eq.~\eqref{eq_RelativisticLoss} encourages the generator to increase the relative realism of its outputs compared to real PET images. 
The second term is designed to estimate how much more realistic a real image is compared to a generated one, rather than assigning a realism score to each image in isolation.
%estimate the probability that a real image is more realistic than a generated one, rather than assessing each image independently.
}  
Compared to conventional GANs, which can suffer from gradient saturation and mode collapse due to the discriminator's binary real/fake decisions, the relativistic mechanism has been shown to provide a smoother learning signal that stabilizes adversarial training.
It can also improve image quality by preventing the discriminator from becoming overly confident in real-vs-fake discrimination, to maintain meaningful gradients throughout training.
Applied jointly with the diffusion loss, it promotes more consistent improvements in the perceptual quality of the diffusion model and guides the model toward producing more realistic outputs.

%\subsubsection{Gradient Penalty.}
%(3) \textbf{\emph{Gradient Penalty}}: 
To facilitate model training, we also incorporate a zero-centered \emph{gradient penalty} on both real and generated data~\cite{huang2024gan}. 
Specifically, we %adopt the R1 and R2 regularizations, which 
penalize the squared norm of the discriminator's gradients with respect to its inputs. 
%The first penalty constrains the discriminator's gradients on real PET images, while the second one  applies the same constraint on generated predictions \( \hat{x}_0 \), 
This penalty constrains the gradient norm of the discriminator $D$ on the real PET image \( x_0 \) and the generated image \( \hat{x}_0 \), which is 
formulated as:
% \begin{align}
% \mathcal{R}_1(D) &= \mathbb{E}_{\substack{x_0 \\ \sim q(x_0)}} \left[ \left\| \nabla_{x_0} D(x_0) \right\|^2 \right], \\
% \mathcal{R}_2(D) &= \mathbb{E}_{\substack{\hat{x}_0 \\ \sim q(\hat{x}_0)}} \left[ \left\| \nabla_{\hat{x}_0} D(\hat{x}_0) \right\|^2 \right],
% \end{align}
\begin{equation}
%\begin{multline}
\small
\begin{split}
\mathcal{L}_{G} 
%=& \mathcal{R}_1(D) + \mathcal{R}_2(D) \\
=\,\,&\mathbb{E}_{\substack{x_0 \sim \mathbb{P}}} \left[ \left\| \nabla_{x_0} D(x_0) \right\|^2 \right] \\
&
+\mathbb{E}_{\substack{\hat{x}_0 \sim \mathbb{Q}}} \left[ \left\| \nabla_{\hat{x}_0} D(\hat{x}_0) \right\|^2 \right],
%\end{multline}
\end{split}
\end{equation}
where $\nabla$ is the gradient operation. 
{\color{black}This gradient penalty (GP) is designed to promote smoother discriminator behavior on both real and synthesized PET images, improving convergence and ensuring stability throughout training~\cite{huang2024gan,nagarajan2017gradient}}.

\subsubsection{Overall Training Objective} 
The RelA-Diffusion framework is trained by minimizing a hybrid objective function that balances the diffusion model's noise prediction accuracy with the image constraint and the gradient-penalized adversarial guidance for realism. 
The hybrid loss %and discriminator loss $\mathcal{L}_{TD}$ 
is formulated as:
% \begin{align}
% \mathcal{L}_{TG} = \mathcal{L}_{diffusion} + \mathcal{L}_{I} + \lambda_{adv} \mathcal{L}_{G}, \\
% \mathcal{L}_{TD} = \lambda_{adv}\mathcal{L}_{D} + \lambda_{gp} \mathcal{L}_{GP},
% \end{align}
\begin{equation}
\mathcal{L} = \mathcal{L}_{N} + \mathcal{L}_{I} + 
%\lambda_{a} 
\mathcal{L}_{R} 
 + 
 %\lambda_{g} 
\mathcal{L}_{G}. 
\label{eq_hybridLoss}
\end{equation}
\if false
where $\lambda_{a}$ and $\lambda_{g}$ are weighting hyperparameters that control the influences of the relativistic adversarial loss and the gradient penalty, respectively.
\fi

\subsection{%Network Architecture
Implementation}
%\textbf{Generator:} 
The \emph{generator} is implemented using a U-Net architecture, consisting of 3 downsampling and 3 upsampling blocks. 
Each downsampling block contains one convolutional layer for spatial reduction and two residual sub-blocks incorporating skip connections. 
Each residual sub-block consists of two convolutional layers, each followed by group normalization and timestep embedding, along with a shortcut connection. 
Positional embeddings for the diffusion timestep $t$ plus tracer label $c$ project encoded features to match the number of feature channels, enabling effective feature integration.
The multi-sequence MRI inputs are concatenated along the channel dimension and then fed into U-Net along with the input $x_t$. 
The \emph{discriminator} is based on a PatchGAN discriminator architecture~\cite{wang2018high}, designed to operate on local image patches rather than the entire image. 
It consists of 3 convolutional layers with decreasing spatial resolution and increasing channel depth. 
Each convolutional layer is followed by LeakyReLU activation (with a negative slope of 0.2) and batch normalization, except for the final output layer. 
The output of the discriminator is a single scalar representing the relative realism of the input.

%\subsection{Implementation Details}
RelA-Diffusion is implemented using  MONAI~\cite{cardoso2022monai} on PyTorch and trained on a computing cluster with multiple NVIDIA H100 GPUs (each with 80GB memory). 
We use the Adam optimizer for both  generator and discriminator. The learning rate for the generator is set to $5 \times 10^{-5}$ and for the discriminator to $5 \times 10^{-6}$. 
Training is performed with a batch size of 3.    
    The diffusion process is set to $T=1000$ timesteps with a linear noise schedule for $\beta_t$ values, ranging from $\beta_1 = 0.0005$ to $\beta_T = 0.0195$.
    %\item     \textbf{Loss Weights:} 
    The model is trained for 100 epochs.
    The adversarial loss $\mathcal{L}_{R} + \mathcal{L}_{G}$ is weighted by 0.1. % throughout all experiments.

%\subsection{Implementation}
\textbf{Training Procedure.} %  (Algorithm~\ref{alg:reladiff})} 
The training procedure of RelA-Diffusion is outlined in Algorithm~\ref{alg:reladiff}. 
The model is trained on paired multi-sequence MRI \( (x_{T_1}, x_{T_2}) \) and multi-tracer PET \( x_0 \) data. 
At each iteration, a timestep \( t \) is sampled, and Gaussian noise is added to \( x_0 \) to obtain a noisy input \( x_t \). 
The generator \( G \) takes \( x_t \), MRI inputs, \( t \), and tracer label \( c \) to predict the noise \( \hat{\epsilon} \), from which the clean estimate \( \hat{x}_0 \) is reconstructed. 
The relativistic adversarial loss uses the activation \( f(x) = -\log(1 + e^{-x}) \).

%\if false
\begin{algorithm}[!htp]
\footnotesize
\caption{Training Procedure for RelA-Diffusion}
\label{alg:reladiff}
\textbf{Input}: Paired data $\{x_{T_1}, x_{T_2}, x_0, c\}_{i=1}^N$ \\
\textbf{Parameter}: Diffusion steps $T$, learning rate $\eta$ \\
\textbf{Output}: Trained generator $G$

\begin{algorithmic}[1]
\STATE Initialize generator $G$, discriminator $D$, and optimizers
\FOR{epoch $=1$ to $N_{\text{epochs}}$}
    \FOR{each batch $(x_{T_1}, x_{T_2}, x_0, c)$}
        \STATE Sample timestep $t \sim \{1, \dots, T\}$ uniformly
        \STATE Sample noise $\epsilon \sim \mathcal{N}(0, \mathbf{I})$
        \STATE Generate noisy PET: $x_t = \sqrt{\bar\alpha_t} x_0 + \sqrt{1 - \bar\alpha_t} \, \epsilon$
        \STATE Concatenate $x_t$, $x_{T_1}$, and $x_{T_2}$ along the channel dimension as model input
        \STATE Embed timestep $t$ and tracer label $c$; inject into $G$ 
        \STATE Predict noise: $\hat{\epsilon} = G(x_t, x_{T_1}, x_{T_2}, t, c)$
        \STATE Compute $\hat{x}_0 = \frac{x_t - \sqrt{1 - \bar\alpha_t} \hat{\epsilon}}{\sqrt{\bar\alpha_t}}$
        \STATE \textbf{Noise loss:} $\mathcal{L}_N = \frac{1}{n} \sum_{i=1}^n (\epsilon - \hat{\epsilon})^2$
        \STATE \textbf{Image loss:} $\mathcal{L}_I = \frac{1}{n} \sum_{i=1}^n |x_0 - \hat{x}_0|$
        \STATE \textbf{Relativistic adversarial loss:} $\mathcal{L}_R = f(D(\hat{x}_0) - D(x_0)) + f(D(x_0) - D(\hat{x}_0))$
        \STATE \textbf{Gradient penalty:} $\mathcal{L}_G = \|\nabla_{x_0} D(x_0)\|^2 + \|\nabla_{\hat{x}_0} D(\hat{x}_0)\|^2$
        \STATE \textbf{Total loss:} $\mathcal{L} = \mathcal{L}_N + \mathcal{L}_I + \mathcal{L}_R + \mathcal{L}_G$
        \STATE Update generator $G$ using $\nabla \mathcal{L}$
        \STATE Update discriminator $D$ using $\nabla \mathcal{L}$
    \ENDFOR
\ENDFOR
\STATE \textbf{return} trained generator $G$
\end{algorithmic}
\end{algorithm}
%\fi 

\textbf{Inference Procedure.}  %(Algorithm~\ref{alg:reladiff_inference})}
Algorithm~\ref{alg:reladiff_inference} describes the inference process for RelA-Diffusion. 
Given a trained generator \( G \), multi-sequence MRI inputs \( x_{T_1} \) and \( x_{T_2} \), and a target tracer label \( c \), the model synthesizes a PET image via an iterative denoising process. 
Starting from a pure Gaussian noise image \( x_T \sim \mathcal{N}(0, \mathbf{I}) \), the model runs the reverse diffusion process for \( T \) steps. 
At each step \( t \), the generator predicts the noise \( \hat{\epsilon} \), and uses it to estimate a cleaner image \( x_{t-1} \). 
This iterative refinement continues until \( x_0 \) is produced.

%\if false
\begin{algorithm}[!tp]
\footnotesize
\caption{Inference with Trained RelA-Diffusion Generator}
\label{alg:reladiff_inference}
\textbf{Input}: Multi-sequence MRI inputs $(x_{T_1}, x_{T_2})$, tracer label $c$, trained generator $G$ \\
\textbf{Parameter}: Diffusion steps $T$, noise schedule $\{\alpha_t, \bar\alpha_t, \sigma_t\}_{t=1}^T$ \\
\textbf{Output}: Synthesized PET image $\hat{x}_0$

\begin{algorithmic}[1]
\STATE Initialize $x_T \sim \mathcal{N}(0, \mathbf{I})$ \hfill \COMMENT{Start from pure Gaussian noise}
\FOR{$t = T$ to $1$}
    \STATE Predict noise: $\hat{\epsilon} = G(x_t, x_{T_1}, x_{T_2}, t, c)$
    % \STATE Estimate clean image: $\hat{x}_0 = \frac{1}{\sqrt{\bar\alpha_t}} (x_t - \sqrt{1 - \bar\alpha_t} \hat{\epsilon})$
    \STATE Sample noise $z \sim \mathcal{N}(0, \mathbf{I})$ if $t > 1$, else $z = 0$
    \STATE Compute denoised image:
    \[
    x_{t-1} = \frac{1}{\sqrt{\alpha_t}} \left( x_t - \frac{1 - \alpha_t}{\sqrt{1 - \bar\alpha_t}} \hat{\epsilon} \right) + \sigma_t z
    \]
\ENDFOR
\STATE \textbf{return} final image $\hat{x}_0$
\end{algorithmic}
\end{algorithm}
%\fi 

\begin{figure*}[t]
\setlength{\abovecaptionskip}{0pt}
\setlength{\belowcaptionskip}{0pt}
\setlength\belowdisplayskip{0pt}
\setlength{\abovecaptionskip}{0pt}
\centering
\includegraphics[width=1\textwidth]{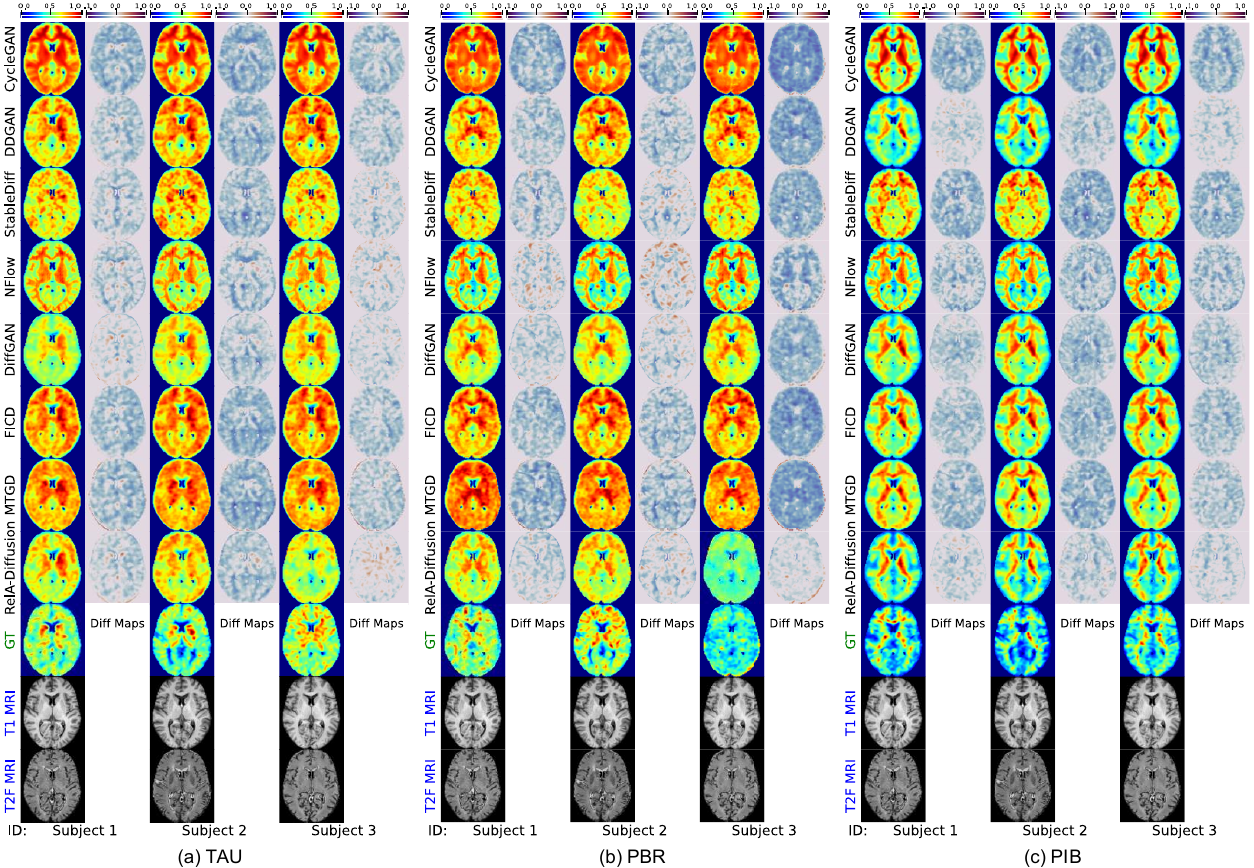}
\caption{Visualization of test PET images for (a) TAU, (b) PBR, and (c) PIB, synthesized by 8 methods, along with corresponding difference (Diff) maps. Ground-truth (GT) and input T1w/T2F MRIs with subject IDs are shown at the bottom.}
\label{fig_qualitative}
\end{figure*}
\section{Experiments}
\label{S_experiment}
% \subsubsection{Materials and Image Pre-processing.} 
% List data and basic pre-processing method.

\begin{table*}[!tbp]
\setlength{\abovecaptionskip}{0pt}
\setlength{\belowcaptionskip}{-8pt}
\setlength{\abovedisplayskip}{0pt}
\setlength\belowdisplayskip{0pt}
\centering
\renewcommand{\arraystretch}{0.7}
\caption{Quantitative results of eight methods for TAU-, PBR- and PIB-PET generation on NFL-LONG (best results in bold). The column \textit{Sig.} indicates statistically significant ($p<0.05$) via $t$-test on SSIM.}
\setlength\tabcolsep{2pt}
%\resizebox{\textwidth}{!}{
\begin{tabular}{@{\extracolsep{\fill}}lccc c| ccc c| ccc c|
@{}}
\toprule
\multicolumn{1}{l|}{\multirow{2}{*}{Method}} 
& \multicolumn{4}{c|}{Synthesized TAU-PET Image}
& \multicolumn{4}{c|}{Synthesized PBR-PET Image}
& \multicolumn{4}{c}{Synthesized PIB-PET Image}\\
\cmidrule(l){2-5} \cmidrule(l){6-9} \cmidrule(l){10-13}
\multicolumn{1}{l|}{} 
& PSNR$\uparrow$ & SSIM$\uparrow$ & MAE$\downarrow$ & \multicolumn{1}{c|}{Sig.} 
& PSNR$\uparrow$ & SSIM$\uparrow$ & MAE$\downarrow$ & \multicolumn{1}{c|}{Sig.}
& PSNR$\uparrow$ & SSIM$\uparrow$ & MAE$\downarrow$ & \multicolumn{1}{c}{Sig.} \\

\midrule

\multicolumn{1}{l|}{CycleGAN~\cite{zhou2021synthesizing}}
& 27.506{\tiny $\pm$1.598} & 0.869{\tiny $\pm$0.013} & \multicolumn{1}{c}{\textbf{0.017}{\tiny $\pm$0.004}} & $*$
& 28.065{\tiny $\pm$2.687} & 0.833{\tiny $\pm$0.022} & \multicolumn{1}{c}{0.017{\tiny $\pm$0.007}} & \multicolumn{1}{c|}{$*$}
& 24.954{\tiny $\pm$2.274} & 0.846{\tiny $\pm$0.017} & \multicolumn{1}{c}{0.024{\tiny $\pm$0.010}} & \multicolumn{1}{c}{$*$} \\

\multicolumn{1}{l|}{DDGAN~\cite{xiao2021tackling}}
& 25.448{\tiny $\pm$2.373} & 0.892{\tiny $\pm$0.015} & \multicolumn{1}{c}{0.025{\tiny $\pm$0.009}} & $*$
& 25.201{\tiny $\pm$3.044} & 0.884{\tiny $\pm$0.017} & \multicolumn{1}{c}{0.027{\tiny $\pm$0.011}} & $*$
& \textbf{26.382}{\tiny $\pm$2.561} & \textbf{0.870}{\tiny $\pm$0.018} & \multicolumn{1}{c}{\textbf{0.021}{\tiny $\pm$0.010}} & \multicolumn{1}{c}{$*$}\\

\multicolumn{1}{l|}{StableDiff~\cite{rombach2022high}}
& 26.483{\tiny $\pm$4.426} & 0.881{\tiny $\pm$0.031} & \multicolumn{1}{c}{0.023{\tiny $\pm$0.015}} & $*$
& \textbf{28.845}{\tiny $\pm$2.681} & 0.892{\tiny $\pm$0.019} & \multicolumn{1}{c}{\textbf{0.015}{\tiny $\pm$0.006}} & $*$
& 25.245{\tiny $\pm$1.735} & 0.852{\tiny $\pm$0.013} & \multicolumn{1}{c}{0.023{\tiny $\pm$0.007}} & \multicolumn{1}{c}{$*$}\\

\multicolumn{1}{l|}{NFlow~\cite{beizaee2023harmonizing}}
& 27.615{\tiny $\pm$3.490} & 0.881{\tiny $\pm$0.020} & \multicolumn{1}{c}{0.019{\tiny $\pm$0.010}} & $*$
& 28.844{\tiny $\pm$2.030} & 0.887{\tiny $\pm$0.013} & \multicolumn{1}{c}{\textbf{0.015}{\tiny $\pm$0.004}} & $*$
& 24.522{\tiny $\pm$1.304} & 0.805{\tiny $\pm$0.017} & \multicolumn{1}{c}{0.024{\tiny $\pm$0.004}} &\multicolumn{1}{c}{$*$}\\

\multicolumn{1}{l|}{DiffGAN~\cite{wang2023diffusiongan}}
& 24.980{\tiny $\pm$2.330} & 0.884{\tiny $\pm$0.013} & \multicolumn{1}{c}{0.026{\tiny $\pm$0.009}} & $*$
& 23.112{\tiny $\pm$2.938} & 0.870{\tiny $\pm$0.021} & \multicolumn{1}{c}{0.034{\tiny $\pm$0.012}} & $*$
& 23.491{\tiny $\pm$2.640} & 0.855{\tiny $\pm$0.021} & \multicolumn{1}{c}{0.031{\tiny $\pm$0.013}} & \multicolumn{1}{c}{$*$}\\

\multicolumn{1}{l|}{FICD~\cite{yu2024functional}}
& 24.236{\tiny $\pm$2.877} & 0.883{\tiny $\pm$0.015} & \multicolumn{1}{c}{0.030{\tiny $\pm$0.012}} & $*$
& 25.594{\tiny $\pm$2.945} & 0.885{\tiny $\pm$0.017} & \multicolumn{1}{c}{0.025{\tiny $\pm$0.010}} & $*$
& 24.229{\tiny $\pm$2.860} & 0.860{\tiny $\pm$0.021} & \multicolumn{1}{c}{0.029{\tiny $\pm$0.012}} & \multicolumn{1}{c}{$*$}\\

\multicolumn{1}{l|}{MTGD~\cite{zhong2025multi}}
& 24.998{\tiny $\pm$1.937} & 0.872{\tiny $\pm$0.012} & \multicolumn{1}{c}{0.026{\tiny $\pm$0.008}} & $*$
& 23.847{\tiny $\pm$2.379} & 0.859{\tiny $\pm$0.012} & \multicolumn{1}{c}{0.031{\tiny $\pm$0.010}} & $*$
& 25.677{\tiny $\pm$2.326} & 0.846{\tiny $\pm$0.013} & \multicolumn{1}{c}{0.023{\tiny $\pm$0.010}} & \multicolumn{1}{c}{$*$}\\

\multicolumn{1}{l|}{RelA-Diffusion~(Ours)}
& \textbf{28.314}{\tiny $\pm$3.392} & \textbf{0.898}{\tiny $\pm$0.017} & \multicolumn{1}{c}{\textbf{0.017}{\tiny $\pm$0.009}} & --
& \textbf{29.324}{\tiny $\pm$2.437} & \textbf{0.898}{\tiny $\pm$0.017} & \multicolumn{1}{c}{\textbf{0.015}{\tiny $\pm$0.006}} & --
& \textbf{26.270}{\tiny $\pm$1.687} & 0.861{\tiny $\pm$0.012} & \multicolumn{1}{c}{\textbf{0.020}{\tiny $\pm$0.006}} & \multicolumn{1}{c}{--}\\

\bottomrule
\label{tab_quantitative1}
\end{tabular}
%}
\end{table*}

\subsection{Materials and Experimental Setup} 

\subsubsection{Materials and Image Pre-processing}
Two datasets are utilized in this study:
(1) NFL-LONG~\cite{walton2022subjective} comprises 152 participants 
with a mean age of $60.15 \pm 6.25$ years. 
While all subjects have T1-weighted (T1w) and T2-FLAIR (T2F) MRI scans, the availability of PET tracers varies across individuals: 142 subjects have PBR-PET, 105 subjects have PIB-PET, and 124 subjects have TAU-PET using the AV1451 tracer.  
A subset of 80 subjects possesses all imaging modalities.
From this complete cohort, 10 subjects are used for testing across all three tracer types, while the remaining data are used for training. 
Multi-tracer PET from the same session is rare in public datasets.
(2) ADNI~\cite{jack2008alzheimer}, a public dataset, contains 502 cognitively normal subjects with available T1w MRI, T2F MRI, and TAU-PET scans (with the AV1451 tracer) in the baseline visit. 
\textcolor{black}{We use all 502 subjects as an external evaluation cohort to assess the generalizability of our model.} 
The mean age of these ADNI subjects is $69.83 \pm 8.80$. 
The subject IDs used are provided in the \emph{Supplementary Materials} for reproducibility. 
T1w MRI scans are registered to the MNI space.
The PET images are rigidly aligned to the corresponding T1w MRI in subject space and then nonlinearly warped to the MNI space using the MRI-derived deformation field. 
T2F images follow the same pipeline as PET, with an additional bias correction applied prior to rigid alignment. 
All volumes are resampled to an isotropic resolution of $1 \times 1 \times 1 \mathrm{mm}^3$, and a standard MNI brain mask is applied for skull removal.
To exclude irrelevant background, all images are center-cropped to $160 \times 180 \times 160$. During training, voxel intensities are scaled to the range $[-1, 1]$, while evaluation uses the normalized range $[0, 1]$ for visual and quantitative analyses.

\subsubsection{Experimental Setup}
We quantitatively evaluate image quality using three metrics: peak signal-to-noise ratio (PSNR), structural similarity index (SSIM), and mean absolute error (MAE). 
Prior to evaluation, all synthesized PET images are intensity-normalized to [0, 1] and padded to match their original spatial dimensions. 
% {\color{red}1) How to set training/validation/test data? 2) If we use SUVr and CenTauR, we need to say something here.} 加到Materials and Image Pre-processing.里了
%\subsubsection{Competing Methods.}  
We compare our method against seven SOTA methods for medical image synthesis. 
% (1) GAN~\cite{pan2021disease}: A generative adversarial network trained with adversarial and $l_1$ losses, producing a four-channel output for multi-tracer PET synthesis. 
(1) \textbf{CycleGAN}~\cite{zhou2021synthesizing}: A conditional GAN that concatenates tracer labels with the input and employs cycle-consistency loss 
alongside adversarial and $l_1$ losses.
% (3) V-GAN~\cite{larsen2016autoencoding}: A variational autoencoder-GAN hybrid, where a VAE serves as the generator, and a discriminator is used for adversarial training. 
(2) Denoising Diffusion GAN (\textbf{DDGAN})~\cite{xiao2021tackling}: This method models each denoising step of a diffusion process using a multi-modal conditional GAN, allowing for faster sampling while maintaining high sample quality and diversity by improving mode coverage and training stability.
% (3) \textbf{Stable Diffusion}~\cite{rombach2022high,wu2023structural}: A latent diffusion model that encodes 3D images into a compact latent space ($10\times12\times10$) before applying diffusion-based modality translation, and then uses a decoder to recover the image from the latent space. {\color{red}The encoder and decoder are pretrained and frozen}. %这个pretrained & frozen很不清楚，最好加上Following~\cite{}, the encoder and decoded and ..... 
(3) Stable Diffusion (\textbf{StableDiff})~\cite{rombach2022high}: A latent diffusion model that encodes 3D images into a compact latent space before applying diffusion-based modality translation, and then uses a decoder to recover the image from the latent space. Its encoder and decoder components are pretrained on all training PET images in this study and kept frozen during the diffusion model's training for the modality translation task.
(4) Normalizing Flow (\textbf{NFlow})~\cite{beizaee2023harmonizing}: A normalizing flow-based model trained end-to-end, leveraging normalizing flows to guide PET synthesis. 
% (5) \textbf{Diffusion-GAN}~\cite{wang2023diffusiongan}: XX. 
(5) Diffusion GAN (\textbf{DiffGAN})~\cite{wang2023diffusiongan}: This GAN framework injects instance noise, generated by a forward diffusion chain, into the discriminator's input. This approach aims to address GAN training instability and mode collapse by providing more stable and diverse learning signals.
(6) \textbf{FICD}~\cite{yu2024functional}: A denoising diffusion probabilistic model that learns a reverse denoising process within a Markov chain and constrained by the clear intermediate prediction. %, {\color{red}conditioned by tracer labels} and MRI inputs. 
%可是FICD论文里并没有tracer label作为input，所以这里说法不准确；会让人误认为这个方法可以利用不同tracer label生成multi-tracer PET？
% (5) Control Net: XX. 
% (7) \textbf{MTGD}~\cite{zhong2025multi}: A diffusion model that uses multi-sequence MRI to generate multi-tracer PET, with the input modalities fused by the cross-attention technique. %那么这个方法和我们用一样的输入吗？需要说明
(7) \textbf{MTGD}~\cite{zhong2025multi}: A diffusion-based model for multi-tracer PET synthesis from multi-sequence MRI (T1w, T2w, T2F), using a Multi-Sequence Attention Encoder and cross-attention fusion. The fused representation, combined with timestep and tracer label, conditions the U-Net via its bottleneck.
To ensure a fair comparison, all methods are implemented on 3D images, with architectures and hyperparameters kept as consistent as possible. 
All diffusion models operate at the image level (i.e., FICD, DDGAN, DiffGAN, %Control Net, 
and MTGD) implement the image constraint $\mathcal{L}_I$ supervision on intermediate clean predictions %since the original DDPM works poorly in 3D MRI-to-PET synthesizing. 
as standard DDPM often fails to capture fine structural details in 3D settings. %请注意，我们论文中不叫l1 loss，叫做image loss L_I
Diffusion models (i.e., DDGAN, StableDiff, DiffGAN, FICD, and MTGD) receive input by concatenating T1w and T2F MRI and noisy PET along the channel dimension, and synthesize tracer-specific PET based on the input label. 
The tracer label is embedded together with the diffusion timestep and injected into the network. 
% The unified network training conditioned on tracer label allows shared representation learning across tracer types.
%问题：所有对比diffusion方法也是同时输入所有tracer-PET进模型，用tracer label做condition？和我们一样的方案？需要说
CycleGAN and NFlow input concatenated T1w and T2F MRI and generate three-channel outputs, each representing a tracer-specific PET image.  
%问题：DD-GAN和DiffusionGAN是GANmodel呢还是diffusion model呢？这种是hybrid的模型
%另外，所有方法的名字在图、表、文字中要一致，比如都用StableDiff，都用NFlow等

% {\color{red}We need to say how these methods can generate multi-tracer PET? separately training for each tracer or others? UNCLEAR NOW}

\if false
\begin{figure*}[!t]
\setlength{\abovecaptionskip}{0pt}
\setlength{\belowcaptionskip}{0pt}
\setlength\belowdisplayskip{0pt}
\setlength{\abovecaptionskip}{0pt}
\centering
\includegraphics[width=1\textwidth]{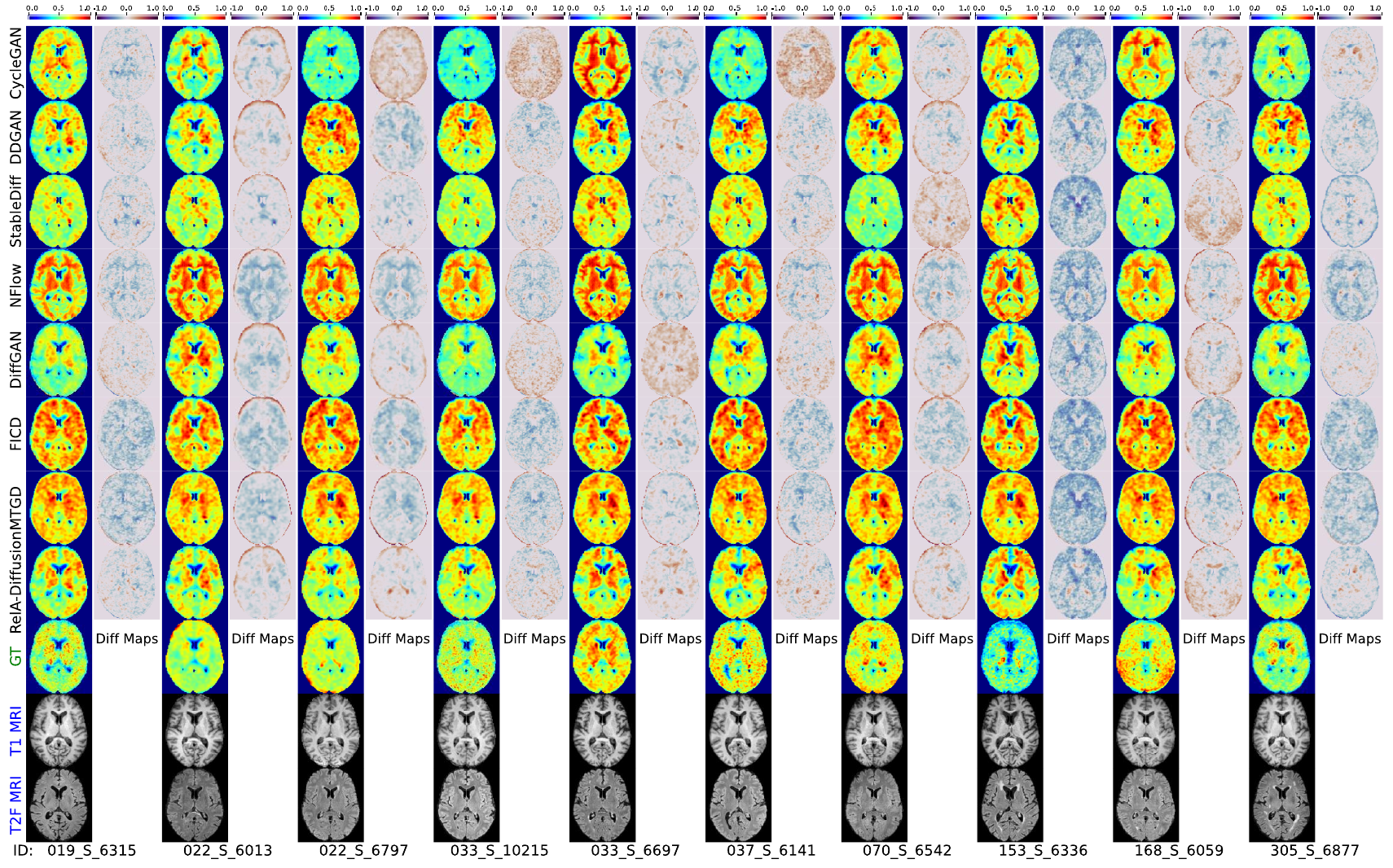}
% \caption{{\color{black}Visualization of 
% test TAU-PET images synthesized by our method and two top competing methods on ADNI}.}
\caption{Visualization of 
test ADNI TAU-PET images synthesized by 8 methods, and difference (Diff) maps. The ground-truth PET images along with the input T1w and T2F MRIs are displayed at the bottom with the corresponding subject IDs.}
\label{tau_only_results}
\end{figure*}
\fi

\begin{figure*}[!t]
\setlength{\abovecaptionskip}{0pt}
\setlength{\belowcaptionskip}{0pt}
\setlength\belowdisplayskip{0pt}
\setlength{\abovecaptionskip}{0pt}
\centering
\includegraphics[width=1\textwidth]{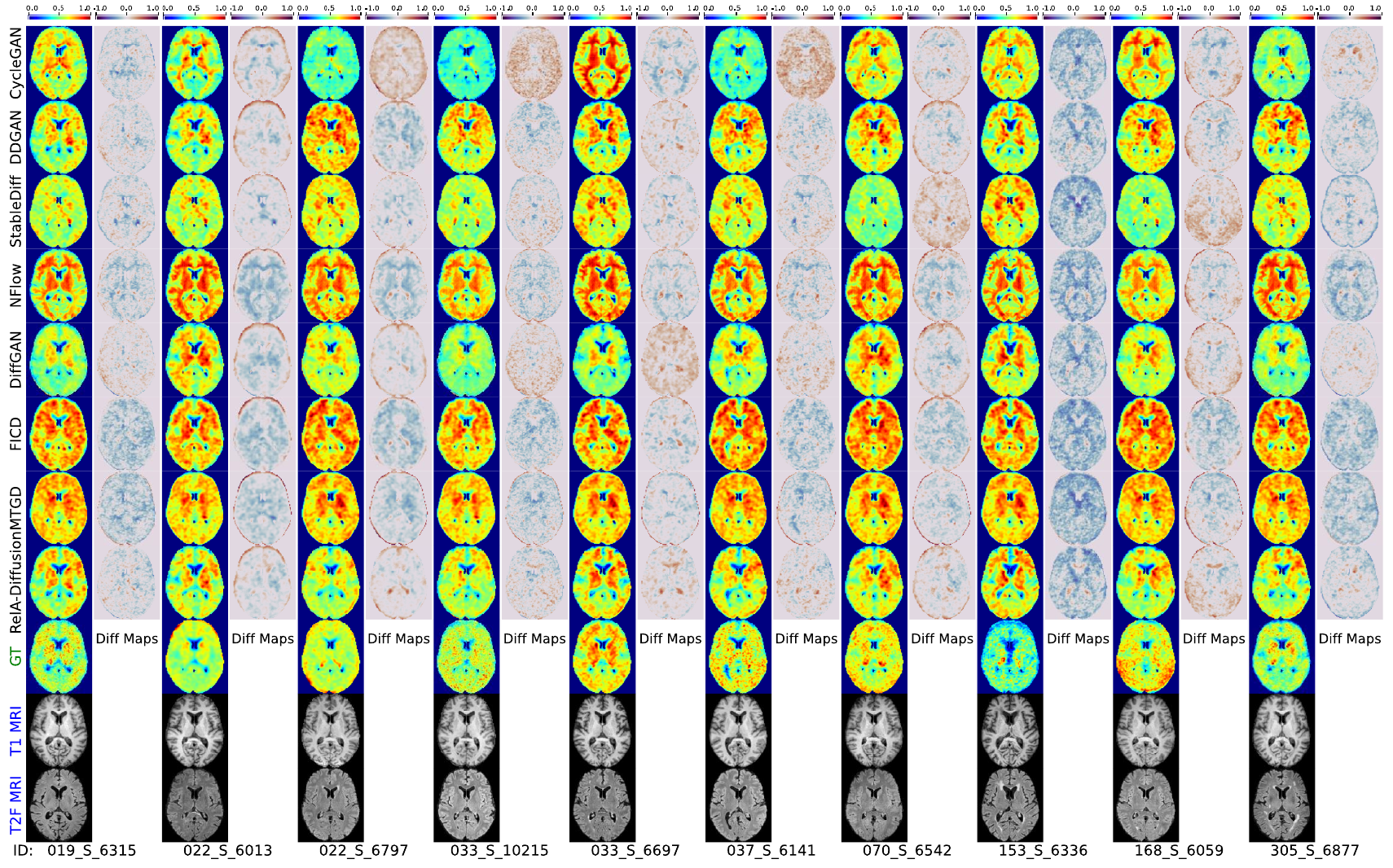}
\caption{Visualization of 
test ADNI TAU-PET images synthesized by 8 methods, and difference (Diff) maps. The ground-truth (GT) PET images along with the input T1w and T2F MRIs are displayed at the bottom with the corresponding subject IDs.}
\label{fig_tau_only_figure}
\end{figure*}

\subsection{Result and Analysis}
\subsubsection{Qualitative Results} 
Figure~\ref{fig_qualitative} displays axial PET slices synthesized by RelA-Diffusion and seven competing methods on the NFL-LONG dataset, alongside the corresponding ground-truth PET images and input T1w and T2F MRIs shown at the bottom. 
For each method, we also visualize difference (Diff) maps between its synthesized PET and the ground truth to highlight voxel-level deviations. 
As shown in Figure~\ref{fig_qualitative}, RelA-Diffusion consistently produces PET images with higher visual quality and closer resemblance to the ground truth compared to the competing approaches. 
This improvement is particularly evident in PBR and PIB tracers, where our method effectively captures subject-specific patterns of neuroinflammation and amyloid deposition in the brain. 
In contrast, the competing methods tend to generate similar outputs across different subjects, failing to reflect individual biological variation. 
Hybrid methods that combine diffusion with adversarial learning (\eg, DiffGAN and RelA-Diffusion) outperform those relying solely on GANs (CycleGAN) or diffusion models (FICD, MTGD, StableDiff). 
%This highlights the benefit of integrating adversarial supervision into the diffusion framework to improve PET image synthesis. Among the hybrid methods, our RelA-Diffusion yields the best visual outputs. 
This demonstrates the advantage of incorporating adversarial supervision into diffusion frameworks for improved PET image synthesis. %Among hybrid approaches, our method produces the best results. 

\subsubsection{Quantitative Results} 
%Table~\ref{tab_quantitative1} reports the quantitative performance of all methods on the NFL-LONG cohort. Across all three PET tracers, RelA-Diffusion consistently achieves the best or highly competitive results. 
Table~\ref{tab_quantitative1} summarizes the quantitative performance of all methods on NFL-LONG. Across all three tracers, RelA-Diffusion consistently achieves the best or highly competitive results. 
For TAU-PET synthesis, it achieves the highest PSNR (28.314) and SSIM (0.898), and ties with CycleGAN for the lowest MAE (0.017), indicating accurate voxel-wise reconstruction and strong %structural 
anatomical preservation. 
For PBR-PET, RelA-Diffusion again achieves the best performance with a PSNR of 29.324, SSIM of 0.898, and MAE of 0.015, outperforming StableDiff and NFlow. 
%While the latter show competitive MAE values, their slightly lower PSNR and SSIM indicate that RelA-Diffusion captures tracer patterns more faithfully. 
Although the latter obtain competitive MAE values, their lower PSNR and SSIM suggest that RelA-Diffusion more faithfully preserves tracer patterns. 
For PIB-PET synthesis, while DDGAN obtains the highest SSIM (0.870), RelA-Diffusion still delivers the best MAE (0.020) and a competitive PSNR (26.270), outperforming other diffusion-based methods (FICD and MTGD). 
% Besides, although CycleGAN achieves a lower MAE for TAU-PET, yet it suffers from relatively low SSIM and PSNR, suggesting that it may generate overly smooth outputs lacking structural fidelity. This observation is consistent with the qualitative results in Figure~\ref{qualitative}, where CycleGAN-synthesized images appear blurred and less capable of capturing fine-grained tracer distribution patterns.
% NFlow performs well for TAU and PBR tracers but significantly underperforms on PIB, likely due to limited modeling capacity under strong distribution shifts. 
% Among prior diffusion models, FICD and MTGD show moderate SSIM but consistently higher MAE and lower PSNR, revealing difficulty in generating anatomically accurate and quantitatively reliable PET images without adversarial regularization.

%\textcolor{red}{t-test}

% \textcolor{red}{
% We conduct paired t-tests between RelA-Diffusion and each baseline across all subjects and metrics. The improvements of RelA-Diffusion are statistically significant ($p < 0.01$) in most comparisons, especially in SSIM and MAE for TAU and PBR, confirming its advantage in synthesizing structurally and quantitatively accurate PET images. These results validate the effectiveness of integrating relativistic adversarial loss and timestep-wise supervision within the conditional diffusion framework.
% }

\subsubsection{Region-Level Evaluation Results}
\begin{figure*}[!t]
\setlength{\abovecaptionskip}{-8pt}
\setlength{\belowcaptionskip}{0pt}
\setlength\belowdisplayskip{0pt}
\setlength{\abovecaptionskip}{0pt}
\centering
\includegraphics[width=1\textwidth]{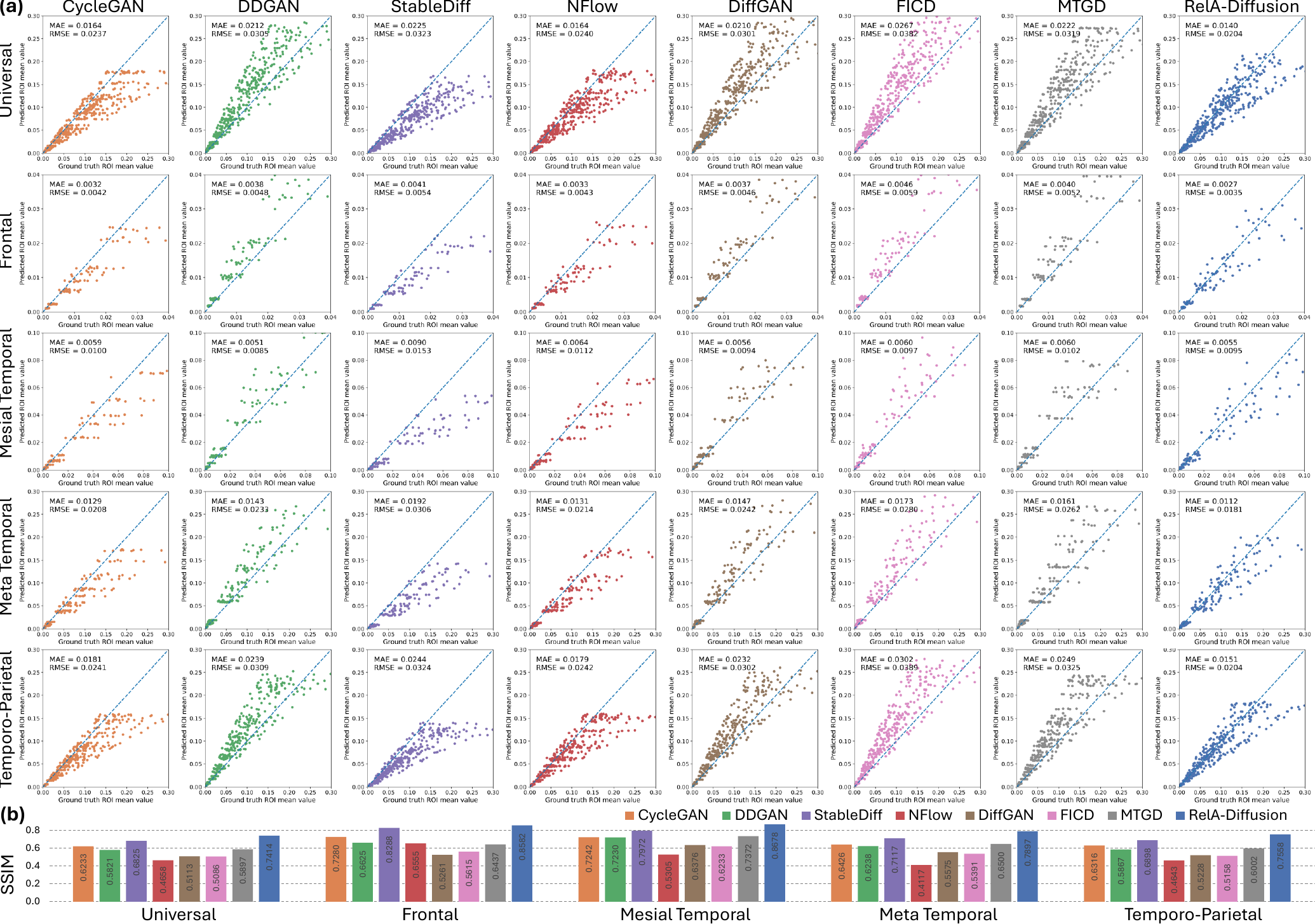}
\caption{Region-level quantitative evaluation on NFL-LONG. (a) Scatter plots of predicted vs. ground-truth volume-of-interest (VOI) mean SUVr for TAU-PET across eight methods and five tau-related VOIs (Frontal, Mesial Temporal, Meta Temporal, Temporo-Parietal, and Universal)~\cite{villemagne2023centaur}, with MAE and RMSE reported in each panel. 
(b) Comparison of SSIM results across the same five VOIs~\cite{villemagne2023centaur}.} %, demonstrating the structural fidelity of RelA-Diffusion across different VOIs.}
\label{tau_roi}
\end{figure*}

{\color{black}
Beyond whole-image analysis, we perform region-level evaluation to assess tracer quantification across five standardized CenTauR volumes-of-interest (VOIs)~\cite{villemagne2023centaur}, comparing synthesized and ground-truth TAU-PET. 
As shown in Figure~\ref{tau_roi}~(a), RelA-Diffusion demonstrates superior consistency between predicted and ground-truth VOI mean SUVr values compared to all competing methods. Our method achieves the tightest clustering along the identity line across all regions, including  Universal, Frontal, Mesial Temporal, Meta Temporal, and Temporo-Parietal VOIs. 
Quantitatively, RelA-Diffusion consistently yields the lowest MAE and root mean square error (RMSE) in every region, including those highly sensitive to early tau accumulation (\eg, Temporal VOIs).  
In Figure~\ref{tau_roi}~(b), we evaluate the structural integrity of synthesized images using masked SSIM across five VOIs. 
RelA-Diffusion demonstrates superior structural fidelity compared to all competing methods, maintaining high consistency particularly in the Frontal (SSIM:~0.8582) and Mesial Temporal (SSIM:~0.8678) regions. 
While competing methods show significant performance degradation in these tau-sensitive areas (with some dropping below 0.50), our framework remains robust across all regions, %including the Universal, Frontal, and Temporo-Parietal VOIs, 
consistently outperforming these SOTA approaches. 
%These results indicate that the synthesized images effectively preserve clinically relevant functional information and fine-grained spatial distribution. 
%This indicates that our relativistic adversarial approach not only recovers the correct mean uptake values but also preserves the fine-grained structural details and spatial distribution of the tracer across diverse brain regions.
%\textcolor{red}{To do:TAU 5 VOIs PIB CTX
}

\section{Discussion}
\label{S_discussion}
In this study, we present RelA-Diffusion, a novel relativistic adversarial diffusion framework designed for high-fidelity multi-tracer PET synthesis from multi-sequence MRI. 
By integrating a gradient-penalized relativistic adversarial loss into the reverse diffusion process, our method effectively balances the preservation of global anatomical structures with the generation of fine-grained, tracer-specific pathological details. 
Our experimental results on the NFL-LONG and the ADNI cohorts demonstrate that RelA-Diffusion significantly outperforms SOTA GAN and diffusion-based approaches in terms of visual fidelity and quantitative accuracy. 
The following sections evaluate the robustness of our model through external generalization, assess its clinical utility in downstream cognitive tasks, and explore the computational efficiency of its latent-space extension. Finally, we provide an ablation study to validate the contribution of each framework component and discuss current limitations alongside directions for future work.

\subsection{Generalization to External Data} 
\subsubsection{Quantitative and Qualitative Evaluation}
{\color{black}To assess the generalization capability of our method, we perform external evaluation on ADNI~\cite{jack2008alzheimer}, which includes paired T1w and T2F MRIs along with TAU-PET scans acquired using the {\color{black}AV1451} tracer. 
Without any fine-tuning, the model trained on NFL-LONG is directly applied to 30 ADNI subjects for TAU-PET synthesis. 
Table~\ref{tab_tau_only_results} and Figure~\ref{fig_tau_only_figure} present the qualitative examples and quantitative metrics for multi-sequence MRI to TAU-PET synthesis on ADNI. 

As shown in Table~\ref{tab_tau_only_results}, our method consistently achieves the best performance across all metrics when compared with the seven SOTA methods, demonstrating superior generalization capability. 
Diffusion-only methods (e.g., FICD and MTGD) cannot yield good results compared to hybrid approaches such as DiffGAN, suggesting limited generalizability to unseen data. 
The qualitative results in Figure~\ref{fig_tau_only_figure} further support this observation, where RelA-Diffusion produces images that closely resemble the ground truth, and the corresponding difference (Diff) maps are visibly lighter. 
%These results demonstrate the strong generalizability of the proposed model when applied to external datas acquired from different scanners and scanning protocols. 
These findings highlight the strong generalizability of our method when applied to external datasets acquired from different scanners and scanning protocols, suggesting its greater potential for clinical deployment across diverse imaging environments.
}% and patient populations.

\begin{table}[!tbp]
\setlength{\abovecaptionskip}{0pt}
\setlength{\belowcaptionskip}{-8pt}
\setlength{\abovedisplayskip}{0pt}
\setlength\belowdisplayskip{0pt}
\renewcommand{\arraystretch}{0.7}
\scriptsize
\centering
\caption{Quantitative results of eight methods for synthesized ADNI TAU-PET images (best results in bold).}%, with best results shown in bold.}
\resizebox{0.46\textwidth}{!}
{
% \tiny
\begin{tabular}{@{\extracolsep{\fill}}lccc@{}}
\toprule
\multicolumn{1}{l|}{Method} & PSNR$\uparrow$ & SSIM$\uparrow$ & \multicolumn{1}{c}{MAE$\downarrow$} \\
\midrule
\multicolumn{1}{l|}{CycleGAN~\cite{zhou2021synthesizing}} & 23.766{\tiny $\pm$3.170} & 0.843{\tiny $\pm$0.034} & 0.030{\tiny $\pm$0.015} \\
\multicolumn{1}{l|}{DDGAN~\cite{xiao2021tackling}}
& 20.563{\tiny $\pm$3.126} & 0.839{\tiny $\pm$0.028} & 0.045{\tiny $\pm$0.018} \\

\multicolumn{1}{l|}{StableDiff~\cite{rombach2022high}}
& 22.179{\tiny $\pm$3.279} & 0.838{\tiny $\pm$0.035} & 0.038{\tiny $\pm$0.018} \\

\multicolumn{1}{l|}{NFlow~\cite{beizaee2023harmonizing}}
& \textbf{25.449}{\tiny $\pm$2.635} & 0.851{\tiny $\pm$0.033} & \textbf{0.024}{\tiny $\pm$0.010} \\

\multicolumn{1}{l|}{DiffGAN~\cite{wang2023diffusiongan}}
& 24.797{\tiny $\pm$2.887} & 0.862{\tiny $\pm$0.029} & 0.026{\tiny $\pm$0.011} \\

\multicolumn{1}{l|}{FICD~\cite{yu2024functional}}
& 18.624{\tiny $\pm$2.934} & 0.831{\tiny $\pm$0.028} & 0.058{\tiny $\pm$0.021} \\

\multicolumn{1}{l|}{MTGD~\cite{zhong2025multi}}
& 17.632{\tiny $\pm$2.374} & 0.818{\tiny $\pm$0.023} & 0.064{\tiny $\pm$0.019} \\

\multicolumn{1}{l|}{RelA-Diffusion~(Ours)}
& 23.781{\tiny $\pm$3.993} & \textbf{0.864}{\tiny $\pm$0.034} & 0.032{\tiny $\pm$0.015} \\

\bottomrule
\label{tab_tau_only_results}
\end{tabular}
}
\end{table}

\begin{table*}[!tbp]
\setlength{\abovecaptionskip}{0pt}
\setlength{\belowcaptionskip}{-2pt}
\setlength{\abovedisplayskip}{0pt}
\setlength\belowdisplayskip{0pt}
\renewcommand{\arraystretch}{0.7}
\setlength{\tabcolsep}{5pt}
\centering
\caption{Quantitative results achieved by RelA-Diffusion, as well as their degraded variants for multi-tracer PET generation on NFL-LONG, with best results shown in bold.}
\resizebox{\textwidth}{!}{
\begin{tabular}{@{\extracolsep{\fill}}lccccccccc@{}}
\toprule
\multicolumn{1}{l|}{\multirow{2}{*}{Method}} & \multicolumn{3}{c|}{Synthesized TAU-PET Image}& \multicolumn{3}{c|}{Synthesized PBR-PET Image}& \multicolumn{3}{c}{Synthesized PIB-PET Image}\\ \cmidrule(l){2-10} 
\multicolumn{1}{l|}{}& PSNR$\uparrow$& SSIM$\uparrow$&\multicolumn{1}{c|}{MAE$\downarrow$}& PSNR$\uparrow$& SSIM$\uparrow$& \multicolumn{1}{c|}{MAE$\downarrow$}& PSNR$\uparrow$& SSIM$\uparrow$& MAE$\downarrow$\\ 
\midrule

\multicolumn{1}{l|}{w/oRA}
& 28.156{\tiny $\pm$4.304} & \textbf{0.908}{\tiny $\pm$0.023} & \multicolumn{1}{c|}{0.019{\tiny $\pm$0.012}}
& 29.311{\tiny $\pm$2.547} & \textbf{0.912}{\tiny $\pm$0.018} & \multicolumn{1}{c|}{0.015{\tiny $\pm$0.006}}
& 25.408{\tiny $\pm$2.182} & \textbf{0.877}{\tiny $\pm$0.010} & \multicolumn{1}{c}{0.022{\tiny $\pm$0.006}}\\

\multicolumn{1}{l|}{w/oGP}
& 18.401{\tiny $\pm$2.780} & 0.780{\tiny $\pm$0.027} & \multicolumn{1}{c|}{0.061{\tiny $\pm$0.021}}
& 20.947{\tiny $\pm$2.732} & 0.786{\tiny $\pm$0.029} & \multicolumn{1}{c|}{0.045{\tiny $\pm$0.013}}
& 16.855{\tiny $\pm$0.953} & 0.764{\tiny $\pm$0.008} & \multicolumn{1}{c}{0.066{\tiny $\pm$0.007}}\\

\multicolumn{1}{l|}{w/oRAGP}& 24.956{\tiny $\pm$4.644}       & 0.877{\tiny $\pm$0.038}& \multicolumn{1}{c|}{0.029{\tiny $\pm$0.016}}& 29.261{\tiny $\pm$4.173}& 0.900{\tiny $\pm$0.029}& \multicolumn{1}{c|}{0.016{\tiny $\pm$0.010}}& 23.721{\tiny $\pm$3.166}       & 0.850{\tiny $\pm$0.020}        & \multicolumn{1}{c}{0.029{\tiny $\pm$0.010}}\\

\multicolumn{1}{l|}{w/oT2F}& 28.109{\tiny $\pm$4.457}& 0.898{\tiny $\pm$0.029}& \multicolumn{1}{c|}{0.019{\tiny $\pm$0.013}}& 29.490{\tiny $\pm$2.380}& 0.905{\tiny $\pm$0.017}& \multicolumn{1}{c|}{\textbf{0.014}{\tiny $\pm$0.005}}& 25.366{\tiny $\pm$1.814}       & 0.863{\tiny $\pm$0.012}        & \multicolumn{1}{c}{0.022{\tiny $\pm$0.005}}\\

\multicolumn{1}{l|}{w/oT1w}& 26.022{\tiny $\pm$4.721}       & 0.887{\tiny $\pm$0.035}& \multicolumn{1}{c|}{0.026{\tiny $\pm$0.016}}& \textbf{29.894}{\tiny $\pm$2.839}& 0.907{\tiny $\pm$0.020}& \multicolumn{1}{c|}{\textbf{0.014}{\tiny $\pm$0.006}}& 23.695{\tiny $\pm$2.748}       & 0.851{\tiny $\pm$0.014}        & \multicolumn{1}{c}{0.029{\tiny $\pm$0.008}}\\
\hline
\multicolumn{1}{l|}{RelA-Diffusion}& \textbf{28.314}{\tiny $\pm$3.392} & 0.898{\tiny $\pm$0.017} & \multicolumn{1}{c|}{\textbf{0.017}{\tiny $\pm$0.009}}& 29.324{\tiny $\pm$2.437} & 0.898{\tiny $\pm$0.017}        & \multicolumn{1}{c|}{0.015{\tiny $\pm$0.006}}& \textbf{26.270}{\tiny $\pm$1.687}       & 0.861{\tiny $\pm$0.012}        & \multicolumn{1}{c}{\textbf{0.020}{\tiny $\pm$0.006}}\\ 
\bottomrule
\label{tab_ablation}
\end{tabular}
}
\end{table*}

\subsubsection{Evaluation on Downstream Tasks}
{\color{black}
To evaluate the clinical utility of synthesized PET images, we conduct downstream regression tasks to predict age, Mini-Mental State Examination (MMSE), and Montreal Cognitive Assessment (MoCA) scores on external ADNI data. We compare two input settings: (1) using MRI only, and (2) using MRI combined with synthesized PET (syn-PET) generated from ADNI TAU-PET images. All regression models share the same ResNet-based architecture~\cite{he2016resnet} and training protocol to ensure a fair comparison. 
%The regression analyses 
Results in Fig.~\ref{fig_regression} reveal that, while structural MRI provides a baseline for cognitive score prediction, integrating synthetic PET leads to a marked reduction in prediction error and tighter correlation coefficients (CC) across age, MMSE, and MoCA tasks. 
This performance gain suggests that RelA-Diffusion effectively captures latent tau-pathology signatures (such as the topographic distribution of neurofibrillary tangles) that are not yet manifested as macroscopic cortical thinning in T1w or T2F sequences. 
Specifically, for MoCA and MMSE prediction tasks, which rely on localized cortical integrity, the syn-PET likely provides a surrogate for metabolic decline that precedes the macroscopic structural atrophy captured by MRI. 
%These results highlight that the synthesized images contain clinically relevant functional information that strengthens the model's utility for objective cognitive assessment.
These results indicate that the synthesized images preserve clinically relevant functional information. %, enhancing the model’s utility for objective cognitive assessment.
}

\subsection{Latent Extension of RelA-Diffusion}
\textcolor{black}{We further develop a latent-space variant of our framework, termed RelA-Diffusion-Latent, to improve computational efficiency. 
Instead of directly operating in the voxel space, both conditional MRI inputs and  target PET images are first encoded into a latent space (10×12×10) using a pretrained frozen autoencoder. 
The diffusion model operates entirely in this latent space. % and predicts the noise added to PET latent .
The diffusion processes and the relativistic adversarial supervision remain the same as RelA-Diffusion.
We apply the image constraint $L_I$ directly on the predicted clean latent $\hat{z}_0$ to enforce accurate reconstruction in the latent domain. 
The gradient-penalized relativistic adversarial constraint is applied on the \emph{decoded} PET image $\hat{x}_0$, obtained by passing $\hat{z}_0$ through the decoder, allowing the discriminator to assess perceptual realism in image space. This hybrid latent–image supervision strategy preserves the computational advantages of latent diffusion while retaining strong adversarial guidance for high-fidelity PET synthesis.
Table \ref{tab_latent} compares RelA-Diffusion and RelA-Diffusion-Latent, showing comparable synthesis performance while the latent version significantly reduces inference time. 
RelA-Diffusion-Latent achieves an inference speed of 0.778 s/vol, which is even faster than the GAN baseline (CycleGAN: 1.47 s/vol), while diffusion-only methods such as FICD remain much slower (134.8 s/vol). 
%This implies that the latent variant offers an excellent efficiency–fidelity trade-off for PET synthesis.
This suggests that the latent variant achieves an excellent balance between efficiency and fidelity. % for PET synthesis.
}

\begin{table*}[!tbp]
\setlength{\abovecaptionskip}{-0pt}
\setlength{\belowcaptionskip}{-0pt}
\setlength{\abovedisplayskip}{0pt}
\setlength\belowdisplayskip{0pt}
\centering
\caption{Quantitative results and inference time of RelA-Diffusion and RelA-Diffusion-Latent for TAU-, PBR-, and PIB-PET generation on NFL-LONG. Inference time is measured in seconds per 3D volume.}
\renewcommand{\arraystretch}{0.7}
\setlength{\tabcolsep}{3pt}
\footnotesize
% \resizebox{\textwidth}{!}{
\begin{tabular}{@{\extracolsep{\fill}}lccccccccc| c@{}}
\toprule
\multicolumn{1}{l|}{\multirow{2}{*}{Method}} 
& \multicolumn{3}{c|}{Synthesized TAU-PET Image}
& \multicolumn{3}{c|}{Synthesized PBR-PET Image}
& \multicolumn{3}{c|}{Synthesized PIB-PET Image}
& \multirow{2}{*}{Inference Time$\downarrow$} \\ 
\cmidrule(l){2-10}
\multicolumn{1}{l|}{} 
& PSNR$\uparrow$ & SSIM$\uparrow$ & \multicolumn{1}{c|}{MAE$\downarrow$}
& PSNR$\uparrow$ & SSIM$\uparrow$ & \multicolumn{1}{c|}{MAE$\downarrow$}
& PSNR$\uparrow$ & SSIM$\uparrow$ & MAE$\downarrow$ 
& \\ 
\midrule

\multicolumn{1}{l|}{RelA-Diffusion}
& 28.314{\tiny $\pm$3.392} & 0.898{\tiny $\pm$0.017} & \multicolumn{1}{c|}{0.017{\tiny $\pm$0.009}}
& 29.324{\tiny $\pm$2.437} & 0.898{\tiny $\pm$0.017} & \multicolumn{1}{c|}{0.015{\tiny $\pm$0.006}}
& 26.270{\tiny $\pm$1.687} & 0.861{\tiny $\pm$0.012} & \multicolumn{1}{c|}{0.020{\tiny $\pm$0.006}}
& 126.778 \\  % <-- replace with your actual time

\multicolumn{1}{l|}{RelA-Diffusion-Latent}
& 27.921{\tiny $\pm$3.902} & 0.900{\tiny $\pm$0.022} & \multicolumn{1}{c|}{0.019{\tiny $\pm$0.011}}
& 28.823{\tiny $\pm$1.395} & 0.901{\tiny $\pm$0.012} & \multicolumn{1}{c|}{0.015{\tiny $\pm$0.004}}
& 24.523{\tiny $\pm$2.076} & 0.865{\tiny $\pm$0.010} & \multicolumn{1}{c|}{0.024{\tiny $\pm$0.005}}
& 0.778 \\  % <-- replace with your actual time

\bottomrule
\label{tab_latent}
\end{tabular}
\end{table*}

\begin{figure*}[!t]
\setlength{\abovecaptionskip}{0pt}
\setlength{\belowcaptionskip}{0pt}
\setlength\belowdisplayskip{0pt}
\setlength{\abovecaptionskip}{0pt}
\centering
\includegraphics[width=1\textwidth]{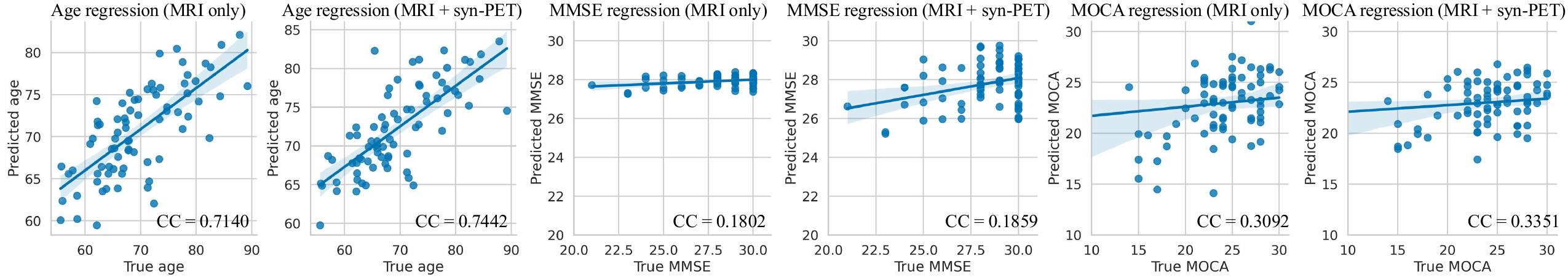}
\caption{Downstream regression performance for age, MMSE, and MoCA prediction using synthesized PET (syn-PET) generated from ADNI TAU images together with MRI, in comparison with MRI-only inputs.}
\label{fig_regression}
\end{figure*}

\subsection{Ablation Study}
{\color{black}
To evaluate the contributions of key components in our RelA-Diffusion framework, we conduct an ablation study on NFL-LONG by comparing it to \textcolor{black}{four} ablated variants: \textcolor{black}{\textbf{(1) w/oRA}, which replaces the relativistic adversarial loss with a standard non-relativistic GAN loss; \textbf{(2) w/oGP}, which removes gradient penalties; \textbf{(3) w/oRAGP}, which replaces the relativistic adversarial loss with a standard non-relativistic GAN loss and removes gradient penalties;} \textbf{(4) w/oT2F}, which excludes T2F input and uses only T1w MRI as condition; and \textbf{(5) w/oT1w}, which uses only T2F MRI and discards T1w input. All variants retain the same network architecture as the full model. % to ensure a fair comparison. 
The quantitative results are reported in Table~\ref{tab_ablation}.
}

{\color{black}{Results in Table~\ref{tab_ablation} demonstrate that removing gradient penalty (GP) leads to the largest performance drop across all tracers, confirming that GP is essential for stabilizing adversarial training and preventing degradation in diffusion–GAN synthesis. Replacing the relativistic adversarial loss (RA) with a standard GAN loss also reduces fidelity, indicating that RA further enhances structural consistency.
When both replacing RA with a standard GAN loss and removing GP (w/oRAGP), there is a consistent drop in PSNR and increased MAE across all tracer types, demonstrating that GP and RA are complementary, and their combination is critical for achieving high-quality PET synthesis. 
Excluding T1w input (w/oT1w) degrades performance, particularly for TAU- and PBR-PET, indicating that the complementary anatomical context captured by T2F and T1w MRIs is critical for accurate tracer distribution modeling. 
Similarly, removing T2F (w/oT2F) also causes a performance drop, but the effect is slightly less pronounced than w/oT1w, suggesting that while both modalities contribute valuable information, T1w may offer more specific pathological contrast for PET synthesis. 
Leveraging both T1w and T2F MRIs with relativistic adversarial supervision, the full model achieves the best performance across all three PET tracers. % on NFL-LONG.
}}

\if false
\subsection{Evaluation on Downstream Tasks}
{\color{black}
To evaluate the clinical utility of synthesized PET images, we conduct downstream regression tasks to predict age, Mini-Mental State Examination (MMSE), and Montreal Cognitive Assessment (MoCA) scores. We compare two input settings: (1) using MRI only, and (2) using MRI combined with synthesized PET (syn-PET) generated from ADNI TAU-PET images. All regression models share the same ResNet-based architecture~\cite{he2016resnet} and training protocol to ensure a fair comparison. 
%The regression analyses 
Results in Fig.~\ref{fig_regression} reveal that, while structural MRI provides a baseline for cognitive score prediction, integrating synthetic PET leads to a marked reduction in prediction error and tighter correlation coefficients (CC) across age, MMSE, and MoCA tasks. 
This performance gain suggests that RelA-Diffusion effectively captures latent tau-pathology signatures (such as the topographic distribution of neurofibrillary tangles) that are not yet manifested as macroscopic cortical thinning in T1w or T2F sequences. 
Specifically, for MoCA and MMSE prediction tasks, which rely on localized cortical integrity, the syn-PET likely provides a surrogate for metabolic decline that precedes the macroscopic structural atrophy captured by MRI. 
%These results highlight that the synthesized images contain clinically relevant functional information that strengthens the model's utility for objective cognitive assessment.
These results indicate that the synthesized images preserve clinically relevant functional information. %, enhancing the model’s utility for objective cognitive assessment.
}
\fi

\subsection{Limitation and Future Work}
%List 2-3 limitations of the current work, and corresponding future research directions.
{\color{black}
Despite the promising results, this study has several limitations. 
\emph{On one hand}, while we evaluated generalizability on ADNI, the extent to which the model is affected by domain shifts, such as variations in PET image characteristics (\eg, noise levels and spatial resolution) caused by different reconstruction algorithms, remains to be fully explored. 
Future work will investigate unsupervised domain adaptation~\cite{ganin2015unsupervised} to enhance robustness across diverse clinical sites. 
\emph{On the other hand}, the iterative nature of voxel-space diffusion models results in high computational costs. 
We aim to explore diffusion distillation techniques%~\cite{huang2023knowledge} 
to accelerate inference. 
%\emph{Finally}, while we utilized T1w and T2F MRI, incorporating other modalities like CT could provide additional pathological context. We plan to extend RelA-Diffusion to handle a broader range of multi-modal inputs for diverse clinical applications.}

\section{Conclusion}
\label{S_conclusion}
{\color{black}This paper presents RelA-Diffusion, a novel relativistic adversarial diffusion framework for synthesizing multi-tracer PET images from multi-sequence MRI. 
By leveraging T1w and T2F MRI inputs, our method captures complementary structural and pathological cues to guide the generation process. 
The proposed gradient-penalized relativistic adversarial loss enhances anatomical realism and stabilizes training. 
Extensive experiments demonstrate that RelA-Diffusion achieves superior performance in visual fidelity and quantitative accuracy compared to existing state-of-the-art methods. 
%This highlights its potential to reduce reliance on expensive, high-risk multi-tracer PET imaging in clinical and research settings. 
%We plan to extend RelA-Diffusion to handle other imaging modalities, such as CT or contrast-enhanced MRI, to support broader clinical applications. We also aim to explore domain adaptation strategies to improve generalizability across diverse scanners.
}

\if false 
\section*{Acknowledgments}
{\color{black}This research was supported in part by NIH grants (Nos. AG073297, AG082938, EB035160, and NS134849). 
A part of the data used in this work is from ADNI and AIBL. 
The ADNI and AIBL investigators provide data but are not involved in data processing, analysis, and writing. 
A comprehensive list of ADNI investigators is accessible \href{https://adni.loni.usc.edu/wp-content/uploads/how\_to\_apply/ADNI\_Acknowledgement\_List.pdf}{online}.
}
\fi

{\color{black}
\footnotesize
\bibliography{bibfile.bib}
\bibliographystyle{IEEEtran}
}

\end{document}